\documentclass[aps,showpacs,twocolumn,pra,citeautoscript,superscriptaddress,floatfix,longbibliography]{revtex4-1}

\usepackage{amsthm}
\usepackage[utf8]{inputenc}
\usepackage{amsmath,amssymb,amsthm,bm,graphicx,xcolor}
\usepackage{natbib}
\usepackage[breaklinks]{hyperref}
\usepackage{color}
\usepackage{tabularx}
\usepackage{epsfig}
\usepackage{graphicx}
\usepackage{wasysym}
\usepackage{dcolumn}
\usepackage{epstopdf}
\usepackage{subfigure}
\usepackage{psfrag}
\usepackage{bbold}
\usepackage[breaklinks]{hyperref}
\hypersetup{colorlinks=true}

\newcommand{\etal}{\emph{et~al.}}

\begin{document}

%\title[title]

\title{Machine Learning the Physical Non-Local Exchange-Correlation Functional of Density-Functional Theory}

\author{Jonathan Schmidt}
\affiliation{Institut f\"ur Physik, Martin-Luther-Universit\"at
Halle-Wittenberg, 06120 Halle (Saale), Germany}
  
\author{Carlos L. Benavides-Riveros}
\email{carlos.benavides-riveros@physik.uni-halle.de}
\affiliation{Institut f\"ur Physik, Martin-Luther-Universit\"at
Halle-Wittenberg, 06120 Halle (Saale), Germany}

\author{Miguel A. L. Marques}
\email{miguel.marques@physik.uni-halle.de}
\affiliation{Institut f\"ur Physik, Martin-Luther-Universit\"at
Halle-Wittenberg, 06120 Halle (Saale), Germany}

\date{\today}

\begin{abstract}
We train a neural network as the universal exchange-correlation
functional of density-functional theory that simultaneously
reproduces both the exact exchange-correlation energy and
potential. This functional is extremely non-local, but retains the
computational scaling of traditional local or semi-local
approximations. It therefore holds the promise of solving some of the
delocalization problems that plague density-functional theory, while
maintaining the computational efficiency that characterizes the
Kohn-Sham equations.  Furthermore, by using automatic differentiation,
a capability present in modern machine-learning frameworks, we impose
the exact mathematical relation between the exchange-correlation
energy and the potential, leading to a fully consistent method. We
demonstrate the feasibility of our approach by looking at
one-dimensional systems with two strongly-correlated electrons, 
where density-functional methods are known to fail, and investigate the
behavior and performance of our functional by varying the degree of non-locality.
\end{abstract}

\maketitle

Nowadays density functional theory (DFT) is the cornerstone of
computational theoretical physics and quantum chemistry, as it
provides the prevalent method for the calculation of the electronic
structure of both solids and molecules. Based on the Hohenberg-Kohn
theorems~\cite{hohenberg1964Inhomogeneous}, DFT reformulates the
quantum many-electron problem as a theory of the ground-state
electronic density $n(\mathbf{r})$. The success of DFT is to a large extent due
to the existence of a system of non-interacting electrons (the
Kohn-Sham system) that has the same ground-state density as the
interacting electrons.  This leads to the Kohn-Sham equations, a set
of self-consistent equations for one-particle
orbitals~\cite{kohnsham}. In such a formalism the ground-sta\-te (GS)
energy can be expressed as:
\begin{equation}
\label{eq:0}
E_{\rm GS}=\sum_i \epsilon_i+E_{\rm xc}[n]-\int\!\!{\rm d}^3 r\; v_{\rm xc}(\mathbf{r}) n(\mathbf{r})
- E_{\rm H}[n],
\end{equation}
where $\epsilon_i$ are the eigenvalues of the Kohn-Sham 
Hamiltonian, $v_{\rm xc}(\mathbf{r})$ is the exchange-cor\-re\-lation potential,
$E_{\rm H}[n]$ is the Hartree energy, and $E_{\rm xc}[n]$ is the 
exchange-cor\-re\-la\-tion energy. The exchange-correlation 
potential is defined as the functional derivative of the universal 
ex\-chan\-ge-cor\-re\-la\-tion energy functional:
\begin{align}
\label{eq:1}
v_{\rm xc}(\mathbf{r}) =\frac{\delta E_{\rm xc}[n]}{\delta n(\mathbf{r})}.
\end{align}
Due to the Hohenberg-Kohn existence theorems, if the exact 
ex\-chan\-ge-correlation energy functional $E_{\rm xc}[n]$ is known,
Kohn-Sham DFT then yields the exact ground-state energy and 
the exact ground-state electronic density. 

Traditionally, ``educated'' formal expressions of the
exchange-correlation energy functional have been proposed by a
combination of theoretical insight, highly accurate
Monte-Car\-lo~\cite{CA} or quantum chemical simulations, or by fitting
general expressions to experimental data. In general, functionals can
be sor\-ted according to Jacob's ladder~\cite{jacobsladder}: the
lowest rung of the ladder is occupied by local-density approximations
(LDA) that use solely single density points as
inputs~\cite{VWN,CP,PW}. The second rung is occupied by
generalized-gradient approximations (GGA) that include the gradient of
the density~\cite{PBE,PBW96}. This is followed by the
meta-GGAs~\cite{Sun2016} (that use the kinetic-energy density) and
hybrid functionals~\cite{doi:10.1063/1.464304,
  doi:10.1063/1.472933,doi:10.1063/1.478522} (that mix a fraction of
non-local Fock exchange) on the subsequent rungs. Note that more than
500 of these functionals have been proposed in the past
decades~\cite{libxc}, although most of them with rather limited
impact.

In spite of the success of DFT in dealing efficiently with electronic
systems, it still suffers from stubborn quantitative and qualitative
failures. For instance, barriers of chemical reactions, band gaps of
materials, or molecular dissociation energies are usually
underestimated \cite{doi:10.1063/1.4869598}.  Degenerate or near-degenerate 
states are also poorly described by DFT.  While hybrid functionals can 
alleviate some of the problems of traditional semi-local functionals, they 
come at a greatly increased computational cost that limits severely the 
number and size of systems that can be researched.  It is believed that 
many of these problems originate in the delocalization and static correlation errors
which plague approximate functionals~\cite{Cohen792,CMSY,PhysRevB.89.195112}. 
Roughly speaking, the delocalization error refers to the tendency of DFT
functionals to spread out the electron density, while the static
correlation arises from the difficulty of describing degenerated
states with a single Slater determinant~\cite{C7CP01137G}.

More recently, machine learning (ML) has revolutionized many fields
of computational sciences, such as image or speech
recognition~\cite{liu2010facial,waibel1990readings}, and has found
countless applications in material science~\cite{butler2019,
  ourreview, PhysRevLett.114.105503}. Within DFT, the application of
ML techniques to the formulation of density functionals has already a
long history~\cite{tozer1996exchange}.  In 2012, an ML approximation
for the kinetic energy functional $T_s[n]$ was constructed for a
system of noninteracting spinless fermions~\cite{snyder2012finding,
  snyder2013orbital}. Yao \etal~\cite{doi:10.1021/acs.jctc.5b01011} developed a 
  convolutional neural network to reproduce the kinetic energy  functional for molecules. They already mentioned the possibility of using the functional derivative of the neural network for minimization purposes.
  In order to exploit the Hohenberg-Kohn
density-potential map, an ML model was later trained to learn the
fundamental relation of DFT between external potentials and electronic
densities~\cite{brockherde2017bypassing}. These works focused mainly
on developing functionals for the total energy or the non-interacting
kinetic energy to facilitate orbital-free DFT calculations.  More
recently, some works have addressed the problem of training the
exchange-correlation potential~\cite{liu2017improving,
  nagai2018neural, PhysRevB.93.235162,PhysRevB.85.235149,
  PhysRevB.94.245129}.  However, this line of research has been
limited by the fact that the exchange-correlation potential was not obtained from the exchange-correlation energy through the functional derivative of Eq.~\eqref{eq:1}.

It is true that one can find in the literature a series of
ap\-proximations to the exchange-correlation functionals that do not
fullfil Eq.~\eqref{eq:1}. For example, the Krieger-Lee-Iafrate
approximation~\cite{KLI} breaks this connection in order to simplify
the implementations of orbital functionals using the optimized effective potential  method~\cite{Sharp_1953,Talman_1976}. Sometimes, it is also
convenient to approximate directly the potential (e.g., in the van
Leeuwen-Baerends GGA from 1994~\cite{LB94} or the modified
Becke-Johnson potential~\cite{BJ,TB09}), leading again to expressions
that do not obey Eq.~\eqref{eq:1}. These so-called ``stray''
functionals~\cite{gaiduk} have found some important applications. For example, the
modified Becke-Johnson is one of the most successful functionals to
calculate electronic band-gaps~\cite{usbenchmark}. Unfortunately, they
are also found to break a series of exact theorems and conditions~\cite{gaiduk},
severely limiting their universality and range of applicability. By
and large, it is highly advantageous to develop consistent functionals
that obey the important Eq.~\eqref{eq:1}.

Modern ML frameworks, like pytorch~\cite{paszke2017automatic} and
tensorflow~\cite{tensorflow2015-whitepaper}, allow for automatic
differentiation with respect to any parameter. Recently, Nagai
\etal\ used this functionality to train exchange-correlation
potentials for molecules~\cite{nagai2019completing}.  They trained
neural networks through a Monte-Carlo updating scheme to reproduce
accurate energies and densities of molecules. The functionals by Nagai
and coauthors follow the traditional approaches of an LDA, GGA,
meta-GGA, and add a related near-region approximation. Although a
clear step forward, us\-ing traditional forms for the
exchange-correlation functional is unlikely to lead to fundamentally
better, disruptive approximations to the exchange-correlation
functionals. New paradigms have to be sought in order to unleash the
power of ML techniques to its full extent.

In this paper we use the auto-differentiation functionality to train
neural-network exchange-correlation functionals through back
propagation. The networks are trained to reproduce not only the
correct exchange-correlation energy $E_{\rm xc}$, but also the
exchange-correlation potential $v_{\rm xc}(r)$ consistently as its
functional derivative with respect to the density.  Consequently, the
resulting functional allows for self-consistent calculations and can
easily be integrated into existing Kohn-Sham DFT
frameworks. Furthermore, these functionals can be made highly
non-local by using the information of the density in a \emph{finite}
neighborhood as input to the neural network, allowing for far more
non-locality than traditional LDA or GGA functionals, despite having
the same computational scaling with system size.  Therefore, this
approach promises to alleviate the delocalization problems of DFT and
to improve its accuracy without the computational expense of hybrid
functionals.  To demonstrate the feasibility of this approach, we
developed an ML functional for the exchange-correlation energy and
exchange-correlation potential based on exact results for two
electrons in one-dimension (1D).

The letter presents the details
of the da\-ta\-set, training process, and neural networks. The exact
dependence of the functional on the degree of locality and its
behavior is also discussed, as well as our
results for the 1D homogeneous electron gas and the H$_2$ molecule along
the dissociation path. Finally, we finish the letter discussing our 
conclusions and fu\-tu\-re research directions.

\textit{Data.}
The training data was produced by solving exactly the one-dimensional
two-electron problem in the external potential generated by up to three
different nuclei. Softening the Cou\-lomb interaction,
\begin{align}
\frac1{r} \rightarrow \frac1{\sqrt{1 + x^2}}, 
\end{align}
we obtain the 1D Hamiltonian driven by the interaction of the two 
electrons, na\-me\-ly, 
\begin{align*}
H(x_1,x_2) &= -\sum_{i=1}^2 \left[\tfrac12 \partial_i^2 +
v(x_i) \right] +
\frac1{\sqrt{1 + (x_1 - x_2)^2}}, 
\end{align*}
where the external potential is given by the superposition of
three potentials, 
\begin{align}
\label{extp}
v(x) = \sum_{k=1}^3 \frac{Z_k}{\sqrt{1 + (x-a_k)^2}}.
\end{align}
The total charge of the nuclei $Z = \sum_k Z_k$ is equal to 2 or 3.
Qualitatively close to real 3D systems,
this 1D model is known as a theoretical laboratory for 
studying strong correlation and developing exchange-correlation
density functionals for DFT \cite{C2CP24118H}.
Since the ground-state problem of the Hamiltonian $H(x_1,x_2)$ 
can be treated as a one-particle problem in two dimensions, the 
problem can be solved exactly. 

We sampled 20\,000 systems and calculated their exact ground-state
energy and ground-state electronic density. We used a grid spacing of
0.1~a.u., and a box size of 20~a.u., leading to a grid with 201 points.
The nuclei positions $a_i$ in Eq.~\eqref{extp} were normally distributed 
with zero mean and variance of 4 a.u. We then solved the corresponding inverse
Kohn-Sham problem in {\sc octopus}~\cite{C5CP00351B} to find the exact
exchange-correlation energy and potential. Since the inversion is
known to be numerically unstable~\cite{Jensen_2018}, we removed
outliers that result from these instabilities. We used up to 12\,800
of these systems for training, 6\,400 for validation during the
training, and 2\,000 systems for the test set. Furthermore, training was
considerably improved when removing outliers with $E_{\rm
  xc}>-0.55$~a.u. from the training set.  No outliers were removed
from the test set to allow for a completely unbiased evaluation of the
functionals. 

In general, one would have to double the data by
mirroring the systems to learn the correct symmetry. However, in this
specific case one can simply build the symmetry directly into the
neural network functional, as explained in the next subsection.
 
\begin{figure}[t]
\includegraphics[width=1.05\linewidth]{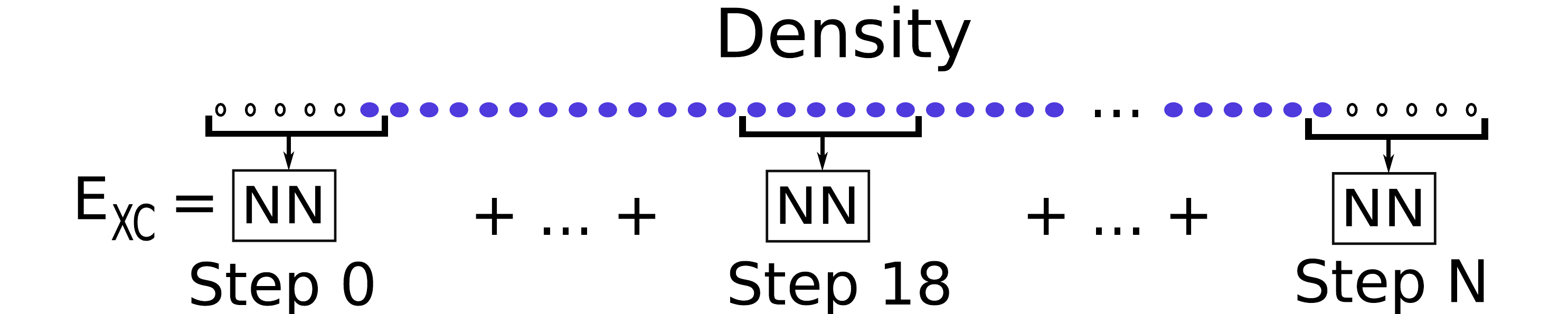}
\caption{Structure of the ML functional in 1D with degree of locality
  equal to $\kappa = 6$ (see text).  At the borders, the density is
  padded with $\kappa -1 = 5$ zeros.  Starting at one of the borders,
  the network calculates the local exchange-correlation energy for
  $\kappa$ points.  In the next step, the input of the network is
  moved by one grid point and it is evaluated again. The network itself is a simple fully connected neural network.}
\label{NNF}
\end{figure}

\textit{Topology of the neural network.}
Our ML functional scans the density of the total system, as
illustrated in Fig.~\ref{NNF}.  The density in a neighborhood of the
test point is used as the input for a 4 or 5 layer fully connected neural network, that then
outputs a \emph{local} exchange-correlation energy. Specifically, the network takes a certain number of density
points as input, which we call $\kappa$, the \textit{kernel
  size}. This is the degree of locality of the ML functional. At the
borders of the system the density is padded with $\kappa -1$ zeros.
Starting at one border the network calculates the local
exchange-correlation energy. In the next step the input of the network
is moved by one grid point, and it is evaluated again. 
As described in Fig.~\ref{NNF}, this process continues until the other 
border is reached.  We arrive at the total exchange-correlation energy
of the system by summing over all network outputs.
The padding and the scanning with a certain kernel-size are inspired by standard convolutional neural networks and can also be implemented as such by concatenating the data along the channel-dimensions in between standard convolutional layers. 
Due to the homogeneity of space, the functional has to be symmetric with respect
to its input densities. The symmetry is ensured by initializing the
weights of the first layer symmetrically along the spacial
dimension. To arrive at the final scalar output we used four- or
five-layer fully connected neural networks.

The possible selection of activation functions (i.e., the non-linearities that follow each multiplication with a weight matrix of a neural network) was rather limited,
because typical functions (e.g.,~rectified linear units) were not
usable due to their lacking differentiability at zero (Using relus actually resulted in piece-wise linear potentials).  To avoid this
problem, we chose exponential linear functions~\cite{ELU}. Different numbers and sizes of hidden layers
were also tested and we settled on the minimum number of parameters
that could be used without underfitting.  In this work all networks
were built on the basis of the ML framework
pytorch~\cite{paszke2017automatic}. The library Ignite~\cite{Ignite}
was used to simplify the training process and tensorboardX to
integrate tensorboard~\cite{tensorflow2015-whitepaper} into pytorch.
The network weights were optimized with Adam~\cite{kingma2014adam}
using default parameters from pytorch.

For the loss function (i.e., the cost function which is going to be 
optimized in the learning process), we should keep in mind that 
the objective is not only to obtain small errors for the 
exchange-correlation energy. To arrive at
the correct density through the solution of the self-consistent Kohn-Sham
equations, also the exchange-co\-rre\-la\-tion potential should be as
close as possible to the exact one. Furthermore, we want not only to
ensure a small error for the potential but also its smoothness. In
addition, the error of the exchange-correlation energy, as well as the
error of the integral $\int\!\!{\rm d}x\: v_{\rm xc}(x) n(x)$ (that
appears in the expression for the total energy~\eqref{eq:0}), should
be minimized.

In order to achieve all these goals concurrently we used the following 
loss function, where $\theta$ are the parameters of the neural network that have to be optimized:
\begin{multline} \label{eq:5}
L(\theta, n_i) = \alpha \text{MSE}(E_{\rm xc})+\beta \text{MSE}(v_{\rm
  xc})\\ +\gamma \text{MSE}\left(\frac{\text{d} v_{\rm
    xc}(x)}{\text{d}x}\right) + \delta \text{MSE}\left(E_{\rm xc}-\int\!\!{\rm d}x\:
v_{\rm xc}(x) n(x) \right).
\end{multline}
This function is a weighted combination of the
mean squared errors (MSE) of the exchange-correlation energy, the
exchange-correlation potential, its numerical spatial derivative and
the difference between the ex\-chan\-ge-correlation energy and the
integral over the potential.  This latter term is part of the formula
for the total energy~\eqref{eq:0} and theoretically allows for some
error cancellation. We also attempted to use the integral as a
separate term in Eq.~\eqref{eq:5}. Depending on the network, one or
the other term produced better results. 
Finally, the weights $\alpha$, $\beta$, $\gamma$ and $\delta$ in 
Eq.~\eqref{eq:5} are also optimized as part of the hyperparameter optimization.
Usual values for $\alpha$, $\beta$, $\gamma$, $\delta$ are 1.0, 100.0, 10.0, 1.0.

The training for the exchange-co\-rre\-lation e\-ner\-gy converges quite 
fast after a few hundred epochs (i.e., one complete pass of the training data). The convergence of the potential 
can take thousands of epochs depending on the training set and 
batch size. At each training step the model was saved if it 
improved the validation error for the potential. The model with the lowest validation error was
later used for the self-consistent Kohn-Sham calculations.
As the amount of memory that is needed per sample is quite limited, 
very large batchsizes (e.g., 4\,096) are possible, allowing for 
a far more efficient parallelization of the training. 
Training with larger batchsizes seems to produce better convergence. 
However, it leads to a strong increase in the error during validation 
with self-consistent Kohn-Sham calculations. Smaller batchsizes 
(32, 64, 128) improve the error by up to 50\% and provide the 
best generalization ability of the functionals, in consistency 
with the literature \cite{masters2018revisiting,Goodfellow-et-al-2016}.

\textit{Evaluation.}
We trained neural networks with various kernel sizes and used them
within a self-consistent Kohn-Sham calculations for a test set of 2\,000
systems created with the method described above.  The
self-consistent Kohn-Sham calculations were run using a self-written
code.  For all kernel sizes, models with different hyper-parameters
were evaluated on a validation set of 250 systems. The training was not completely converged at this
stage. Only for the best models of each kernel size we continued the training
and evaluated the models on the test set.  To compare
various models with respect to the LDA functional of DFT, we chose
energy differences, as these are physically more meaningful.

\begin{table}[t]
\begin{tabular}{ c c c}
 Kernel-size&  MAE(ML)/MAE(LDA)[\%]\\\hline
 LDA &100 \\
 1&38.1\\
 15 &21.8\\
 30 &8.2\\
 60 &8.2\\
 120 &7.1\\
 180 &6.5\\
\end{tabular}
\caption{Mean absolute errors (MAE) for the total energy in self-consistent calculations for various kernel sizes of our ML-DFT functional, relative to the error of the one-dimensional LDA of
  Ref.~\onlinecite{PhysRevA.83.032503}. For reference, the mean absolute error of the LDA of Ref.~\onlinecite{PhysRevA.83.032503} is 1.4$\times10^{-2}$~a.u.}
\label{tablelocality}
\end{table}

\begin{figure} \centering
  \includegraphics[width=0.8\linewidth]{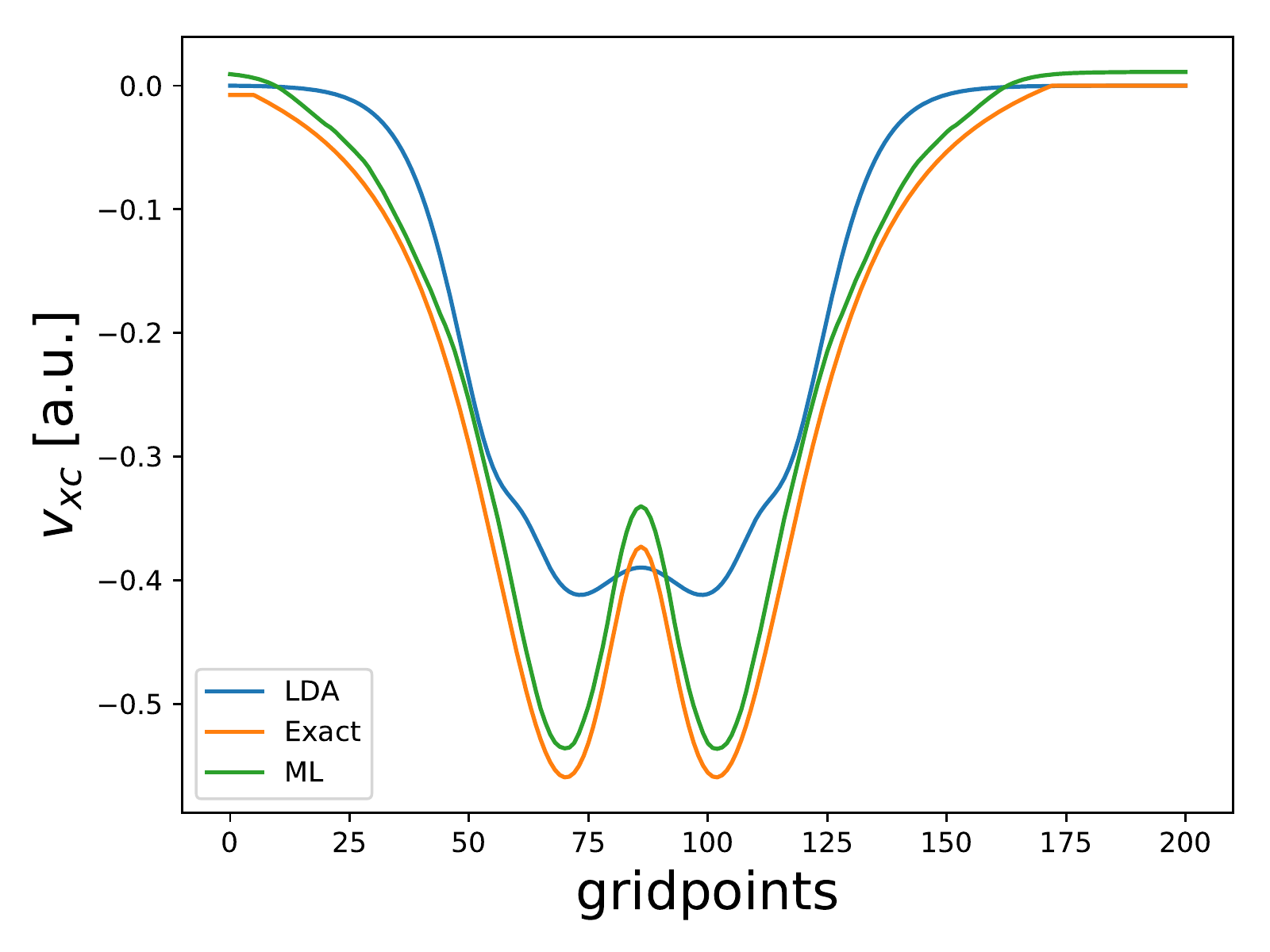}
  \includegraphics[width=0.8\linewidth]{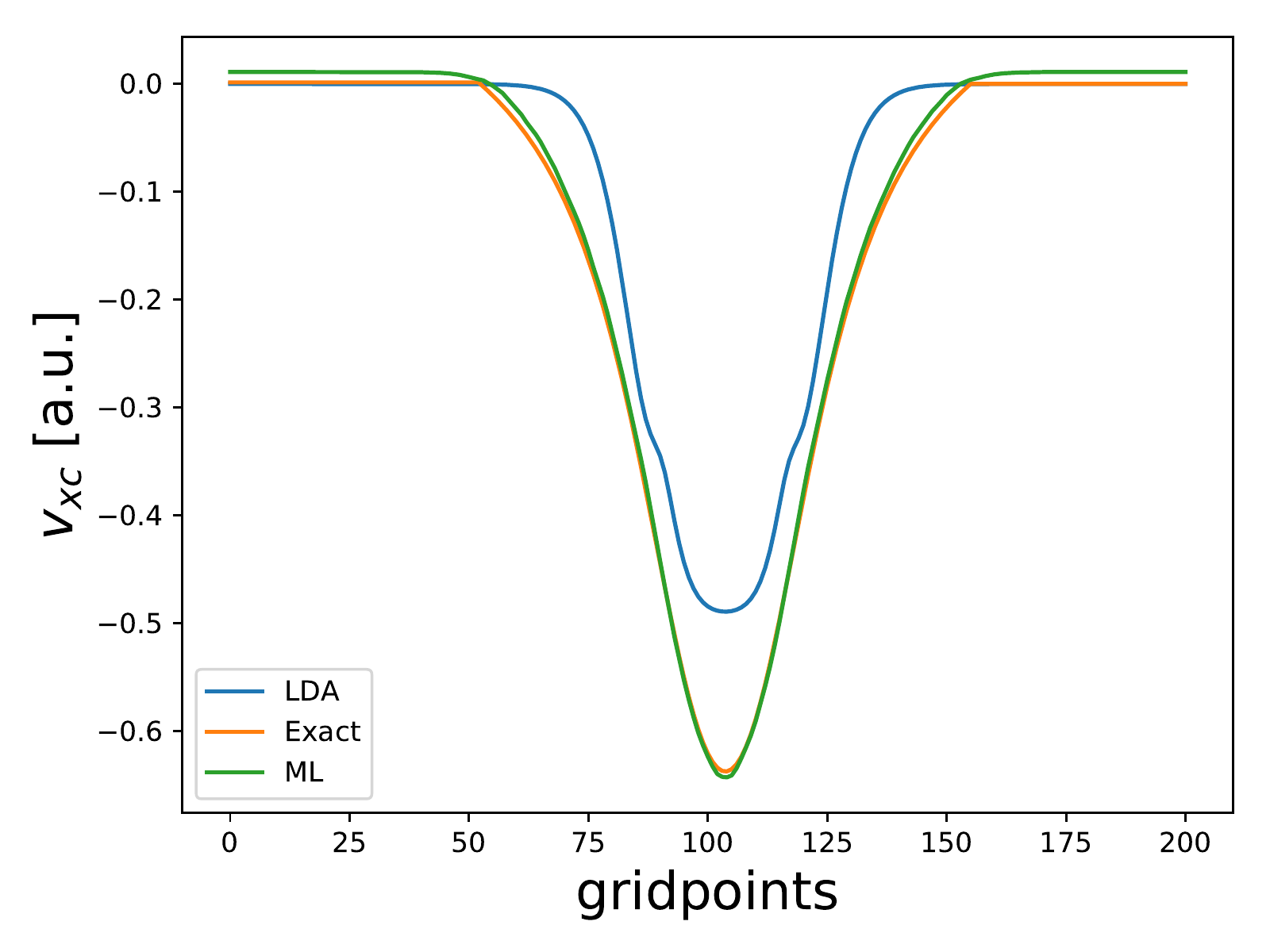}
  \caption{Comparison of exchange correlation potentials of an 1D-LDA \cite{PhysRevA.83.032503}, the exact potential and our ML-functional with kernel-size 30 for two different systems.}
  \label{fig:potential1}
\end{figure}

\begin{figure}
  \centering
  \includegraphics[width=0.8\linewidth]{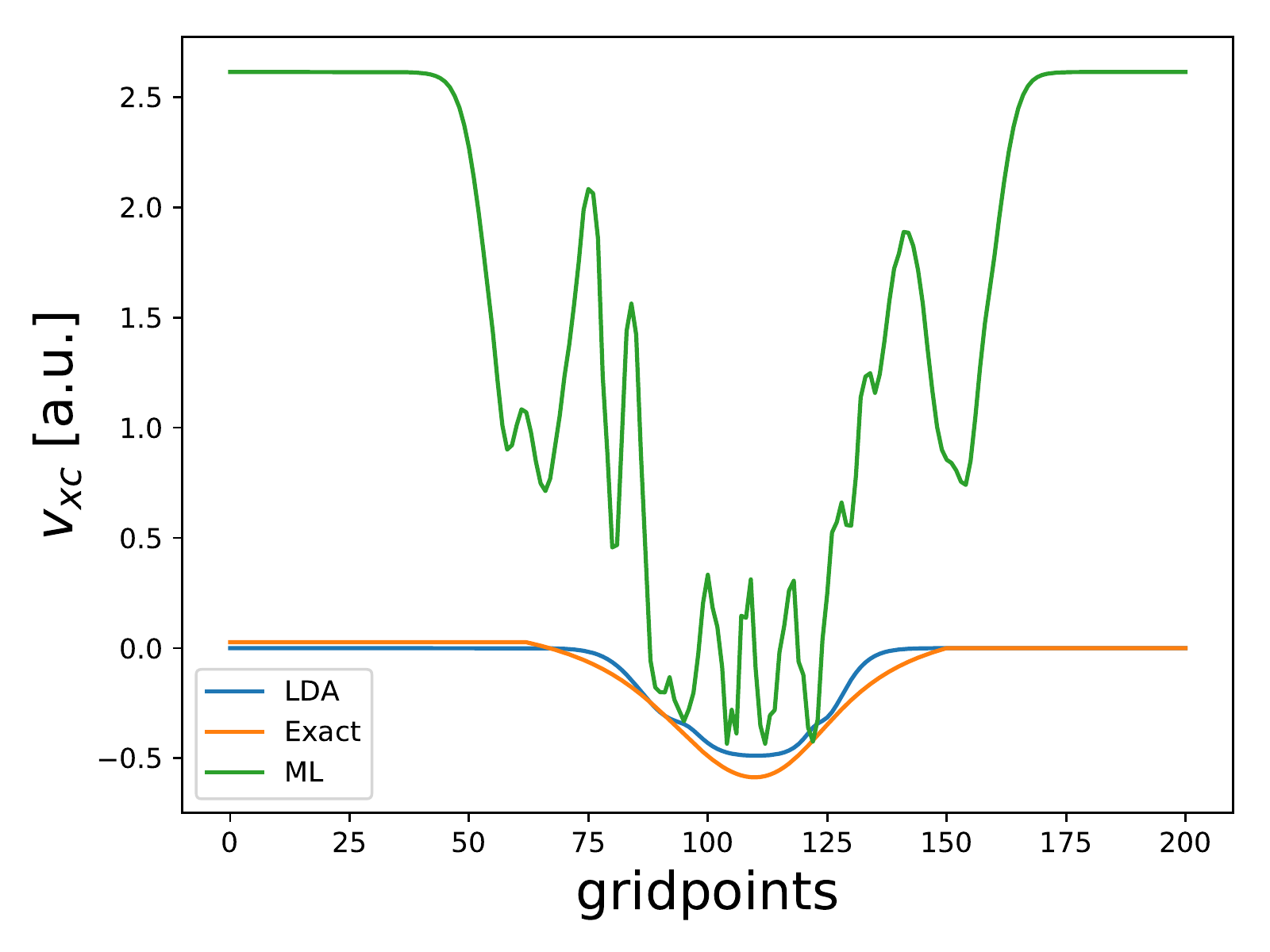}
  \caption{Comparison of exchange correlation potentials of an 1D-LDA \cite{PhysRevA.83.032503}, the exact potential and our ML-functional trained only with the exchange-correlation energy.}
  \label{fig:potential2}
\end{figure}

In Fig.~\ref{fig:potential1}  we plot the exchange-correlation potential resulting from self-consistent calculations with a ML functional trained according to the loss function \eqref{eq:5} for two test-systems and compared it with the exact and LDA predicted exchange-correlation potentials. In Fig.~\ref{fig:potential2}  we plot the same information for one test-system; the ML-functional is this time trained only on the exchange-co\-rre\-lation energy (i.e., not on the potential). Whenever the machine-learned functional is trained both on the energy and the potential, the exchange-correlation potential presents a great improvement in comparison to the traditional LDA functional, while the functional trained only on the energy fails completely. Remarkably, the functionals trained with the loss-function \eqref{eq:5} also show a qualitatively closer behavior to the exact exchange-co\-rre\-lation potentials.
The results for the predicted total energies of the test set
(relative to the energy of H$_2$ at its equilibrium distance) are
presented in Table~\ref{tablelocality}. Our ML-LDA (i.e., the
functional with kernel size $\kappa=1$) already performs better than
the traditional LDA.  As the ML-LDA was trained to reproduce the
exchange-correlation energy of heterogeneous systems while traditional
LDAs are ``trained'' for constant densities the difference in
performance is not surprising.  Increasing the kernel size leads to a
monotonical decrease of the error, and improves the results by more
than a factor of six for the larger sizes. The optimal kernel size is,
in our opinion, around 30 (i.e., 3~a.u.), as larger kernels do not
provide a significant advantage. Furthermore, some of the functionals
with larger kernel sizes also demonstrate unphysical behavior (see
below).

Ultimately, the more non-local the functional is, the higher the
complexity and the larger the number of parameters. This reason, together
with the need to represent more long-range interactions
that are based on different physical principles (such as van der Waals
interaction), makes the training considerably more difficult. One approach to circumvent this problem is to keep
the non-locality limited to ranges on the scale of molecular
bonds. This allows for simpler training and still includes most of
the non-locality that is required for the exchange-correlation energy.
Another possibility would be to enlarge the non-locality by increasing 
the architectural complexity of the functional.

Efforts to decrease the number of training systems for a kernel-size of 30 lead to a slightly increased error of 11\% using 800 samples for training.
Although the scaling of the networks to realistic three-dimensional systems is non-trivial,
we expect that both the number of trainable parameters and 
the density points in the training set will grow cubically when 
transitioning to three-dimensional systems. In this sense, we expect 
a similar demand for training data as in 1D. Furthermore, 
realistic systems are usually far larger and therefore provide 
more ``local'' training samples per system for the neural network. 
Recent research by Nagai and coauthors points in the same direction
\cite{nagai2019completing}.
Indeed, they only requi\-red a few sample molecules and used far more parameters to learn a much more local (and in this sense simpler) functional than the ones used here.

\begin{figure}[t]
\includegraphics[width=1.05\linewidth]{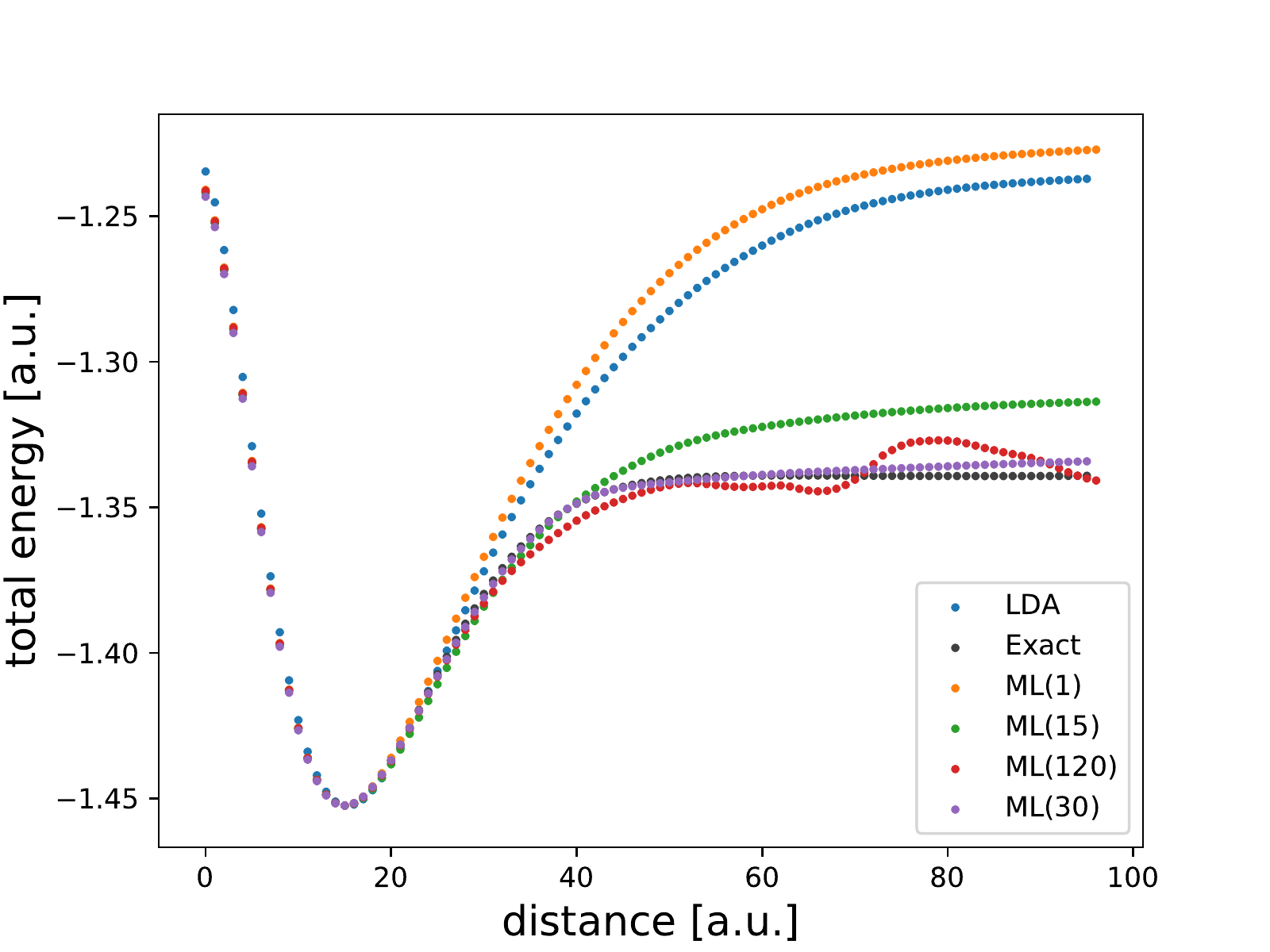}
\caption{Dissociation curves of the 1D H$_2$-molecule 
with ML functionals of varying non-locality (i.e., kernels of 1, 15, 120, 30) 
in comparison to the exact and the LDA results.}
\label{H2}
\end{figure}

Previously, we tested the ML-functional on sample systems belonging to the distribution of the training data.
Now, we go a step
further and test how our functionals perform self-consistently on systems outside this distribution as well as a couple of paradigmatic cases.
The systems in the training set had external potentials arising from 2 and 3 nuclei. In order to go beyond these systems we tested the functional with kernel-size 30 also on a test set of 150 systems with 4 nuclei. Using the same functional as in Table I we arrive at an error for the total energy more than eight times smaller than with the LDA (MAE(ML)/MAE(LDA)=11.9\%). Naturally, the error increased outside the training distribution. However, considering the different nature of the highly charged systems with 4 nuclei, this hints at a good generalization ability of the functional.

In Fig.~\ref{H2} we present our calculations for the H$_2$ mo\-le\-cule in 1D
along the dissociation path with functionals of varying non-locality in 
comparison to the exact result. The curves in Fig.~\ref{H2} are 
shifted to have the same equilibrium energies. As is well-known, the 
traditional LDA completely fails to produce the correct dissociation 
limit~\cite{PhysRevA.83.032503}. The same behaviour is observed for
the ML LDA (with a kernel size of 1). Remarkably, using increasingly more non-local functionals, we can reduce the relative energy error to $3.2\%$ of the LDA error.
It is obvious that even the functional with a kernel size of 
30 will start failing above a certain distance. This is a 
conceptual problem of local KS-DFT and can only be alleviated 
and not eliminated in our approach. It can already be considered a success that our functionals are able to reproduce the dissociation curve 
reasonably well far beyond their own degree of non-locality.
Yet it has to be noted that not all functionals performed that well. 
Some of the functionals with larger kernels 
failed to reproduce a physical behavior with respect to the 
dissociation distance and produced multiple local minima and maxima. 
Despite these problems, they still return the correct equilibrium distance 
and on average far better energies than the LDA.
This unphysical behavior of some functionals just stresses the fact that a rigorous 
validation on a multitude of different systems will 
be essential to arrive at a working functional. It has to be noted that 
a larger training set and a longer training time was far more beneficial 
for this validation than for example the average error. As systems similar 
to the dissociated molecule are most likely outliers of the training data 
this is not surprising. Note that the need for more training data can however be 
avoided by active learning and a thoughtful construction of the training set.

\begin{figure}[t]
\includegraphics[width=1.0\linewidth]{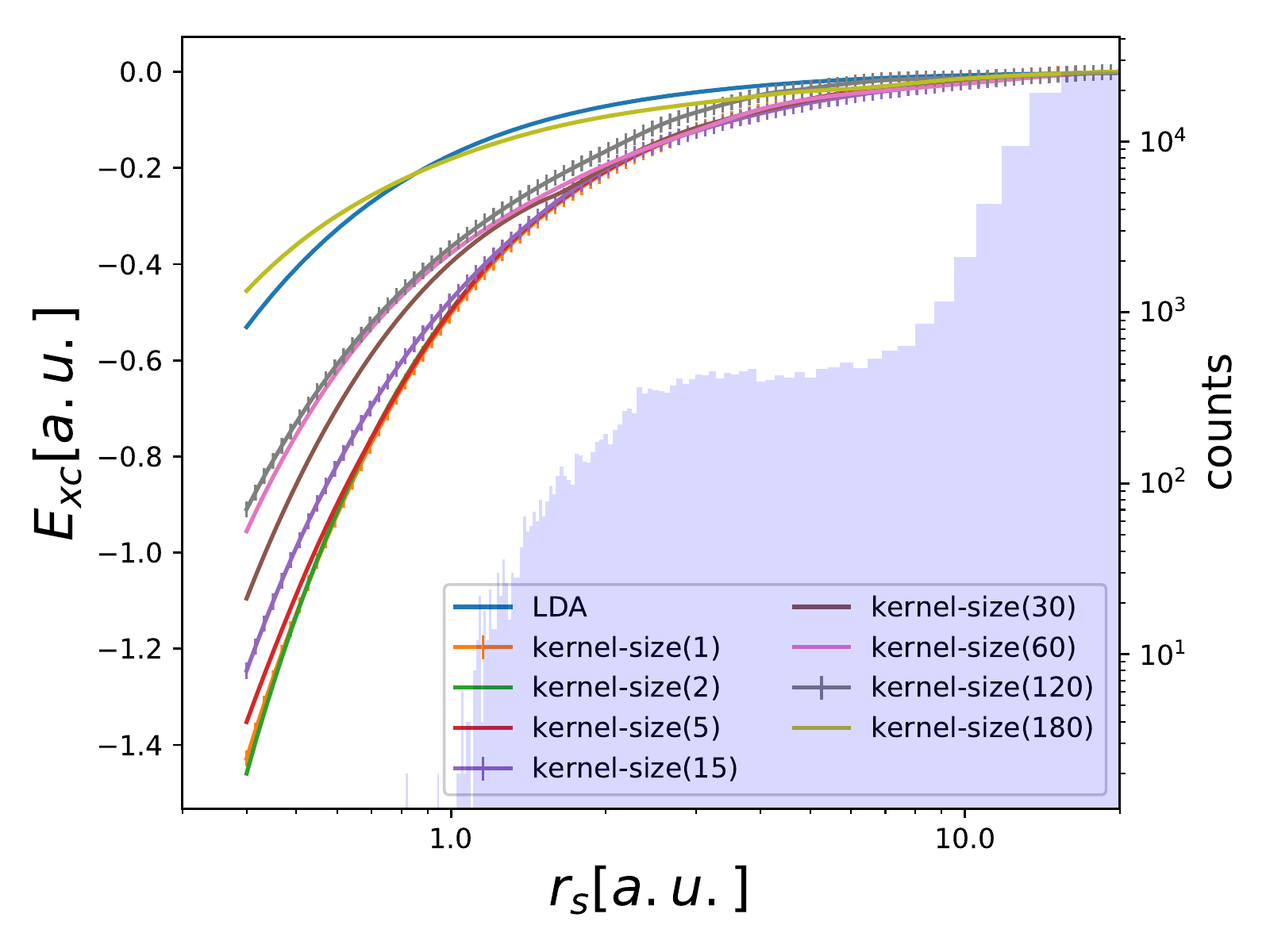}
\caption{The exchange-correlation energy per unit volume of a 
1D homogeneous electron gas from quantum Monte Carlo 
calculations~\cite{PhysRevA.83.032503} (curve labeled LDA) is 
compared with several ML functionals evaluated at constant density.
We also plot a histogram of the number of systems of the training 
set containing more than three grid points with a density
within a bin size of 0.01 a.u. The number of counts can be read 
on the right axis.  Notice that there is basically no system with $r_s < 1$.
The machine-learned curves are shifted to be exactly zero at zero density.}
\label{LDA}
\end{figure}

Finally, we study the homogeneous-electron gas, a model system that is used in the construction of the majority of exchange-correlation functionals. We can simulate this system with our neural networks by providing them with a constant electronic density as input. The results can then be compared to the numerically exact values for the energy density as obtained, for example, from quantum Monte-Carlo simulations~\cite{PhysRevA.83.032503}. Our results are depicted in Fig.~\ref{LDA}  
as a function of the Wigner-Seitz radius $r_s=1/2n$.
Some difficulties for our neural-network functionals are evident. First, the 
functionals were trained only for systems with a specific size while 
the homogeneous electron gas is, in fact, an infinite periodic system. 
Second, as the histogram in 
Fig.~\ref{LDA} illustrates, the training data does not contain almost any 
samples with high densities ($r_s < 1$). Naturally, the availability of training data
similar to the homogeneous electron gas is even more important for the more 
non-local functionals as they take into account larger regions of space.
The first challenge results in the fact that the non-zero biases in 
each layer cause the neural networks to output a non-zero value for 
zero density. When training for different system sizes there are 
several ways to avoid this failure. First one could solve the problem 
by adding systems padded with different amounts of zeros at the border 
to force the neural network to learn the correct relationship. As a 
second possibility, one could force all biases of all layers to zero, 
however, this would severely limit the expressibility of the networks.
To circumvent this problem, and in order to compare the behavior of the 
energy with respect to the Wigner-Seitz radius, we shifted the curves in Fig.~\ref{LDA} to yield zero 
energy for zero density.
Despite the small amount of training data at high density, functionals with larger 
kernel sizes still generalize on average far better to the homogeneous 
electron gas. While this is the case for most models, there are some rare cases, similar to the problems with the H2 dissociation, where large kernel-sizes produce unphysical behavior.

It is not obvious whether this result will remain true in three dimensions, it is nevertheless promising that 
the extra non-local information in the larger kernels might help the 
functionals to be generalized. Constant densities will be an essential 
feature of a functional for solid state physics. Fortunately, exact 
training data in the form of quantum Monte-Carlo calculations already 
exists for this purpose, and can be easily incorporated in our training sets.

\textit{Conclusions.}
In this letter we have demonstrated the viability of learning an 
exchange-correlation potential, via the differentiation of 
the exchange-correlation energy in a physically consistent manner.
This procedure allows for standard self-consistent Kohn-Sham 
calculations. From the presented data, it is evident that 
neural-network functionals trained on the exchange-correlation 
potential and energy have the potential to be far more precise 
than previous local DFT functionals. Increasing the non-locality 
of the functional allows for an extremely precise treatment of 
the electronic interaction on the scale of at least a 
few atomic units and, to a certain extent, even solve long-standing 
problems in DFT like e.g.~molecular dissociation.

For simplicity, we trained a neural network to the one-dimensional two-electron problem 
in the external potential 
generated by up to to three nuclei. Training three-dimensional systems will have 
to be accomplished by using data obtained with coupled-cluster, full configuration-interaction, or quantum Monte Carlo. 
While sufficient data to train a universal functional still has 
to be created, exchange-correlation energies and potentials for a few small molecules already 
exists and can provide a good starting point. The density representation 
on a grid is unfortunately not feasible for more general systems, as grid sizes 
and forms will vary. However, we think this can easily be 
circumvented by representing the density locally in some basis sets
(e.g., Gaussians).

Finally, there is already a long history within
DFT in the development of empirical functionals \cite{PhysRevB.85.235149, PhysRevB.93.235162, 
doi:10.1021/acs.jctc.5b01082, C6SC00705H, Mardirossian2016}. The 
machine learning paradigm allows us to drastically increase the amount 
of data used for the training and the complexity of these functionals. Including known 
exact conditions of the exchange-correlation functional in the 
learning process as constraints in the minimization will still 
be helpful \cite{Hollingsworth2018} and provide further conditions for validation. Furthermore, as the functionals will have to work in 
practically every density environment, the importance of an 
extremely in-depth validation cannot be overstated and will be 
essential to arrive at a widely used functional. 

\textit{Acknowledgments.}
We acknowledge partial support from the German DFG through project
MA-6786/1.

\bibliography{bibliography}

%merlin.mbs apsrev4-1.bst 2010-07-25 4.21a (PWD, AO, DPC) hacked
%Control: key (0)
%Control: author (0) dotless jnrlst
%Control: editor formatted (1) identically to author
%Control: production of article title (0) allowed
%Control: page (1) range
%Control: year (0) verbatim
%Control: production of eprint (0) enabled
\begin{thebibliography}{58}%
\makeatletter
\providecommand \@ifxundefined [1]{%
 \@ifx{#1\undefined}
}%
\providecommand \@ifnum [1]{%
 \ifnum #1\expandafter \@firstoftwo
 \else \expandafter \@secondoftwo
 \fi
}%
\providecommand \@ifx [1]{%
 \ifx #1\expandafter \@firstoftwo
 \else \expandafter \@secondoftwo
 \fi
}%
\providecommand \natexlab [1]{#1}%
\providecommand \enquote  [1]{``#1''}%
\providecommand \bibnamefont  [1]{#1}%
\providecommand \bibfnamefont [1]{#1}%
\providecommand \citenamefont [1]{#1}%
\providecommand \href@noop [0]{\@secondoftwo}%
\providecommand \href [0]{\begingroup \@sanitize@url \@href}%
\providecommand \@href[1]{\@@startlink{#1}\@@href}%
\providecommand \@@href[1]{\endgroup#1\@@endlink}%
\providecommand \@sanitize@url [0]{\catcode `\\12\catcode `\$12\catcode
  `\&12\catcode `\#12\catcode `\^12\catcode `\_12\catcode `\%12\relax}%
\providecommand \@@startlink[1]{}%
\providecommand \@@endlink[0]{}%
\providecommand \url  [0]{\begingroup\@sanitize@url \@url }%
\providecommand \@url [1]{\endgroup\@href {#1}{\urlprefix }}%
\providecommand \urlprefix  [0]{URL }%
\providecommand \Eprint [0]{\href }%
\providecommand \doibase [0]{http://dx.doi.org/}%
\providecommand \selectlanguage [0]{\@gobble}%
\providecommand \bibinfo  [0]{\@secondoftwo}%
\providecommand \bibfield  [0]{\@secondoftwo}%
\providecommand \translation [1]{[#1]}%
\providecommand \BibitemOpen [0]{}%
\providecommand \bibitemStop [0]{}%
\providecommand \bibitemNoStop [0]{.\EOS\space}%
\providecommand \EOS [0]{\spacefactor3000\relax}%
\providecommand \BibitemShut  [1]{\csname bibitem#1\endcsname}%
\let\auto@bib@innerbib\@empty
%</preamble>
\bibitem [{\citenamefont {Hohenberg}\ and\ \citenamefont
  {Kohn}(1964)}]{hohenberg1964Inhomogeneous}%
  \BibitemOpen
  \bibfield  {author} {\bibinfo {author} {\bibfnamefont {P.}~\bibnamefont
  {Hohenberg}}\ and\ \bibinfo {author} {\bibfnamefont {W.}~\bibnamefont
  {Kohn}},\ }\bibfield  {title} {\enquote {\bibinfo {title} {Inhomogeneous
  electron gas},}\ }\href {\doibase 10.1103/physrev.136.b864} {\bibfield
  {journal} {\bibinfo  {journal} {Phys. Rev.}\ }\textbf {\bibinfo {volume}
  {136}},\ \bibinfo {pages} {B864} (\bibinfo {year} {1964})}\BibitemShut
  {NoStop}%
\bibitem [{\citenamefont {Kohn}\ and\ \citenamefont {Sham}(1965)}]{kohnsham}%
  \BibitemOpen
  \bibfield  {author} {\bibinfo {author} {\bibfnamefont {W.}~\bibnamefont
  {Kohn}}\ and\ \bibinfo {author} {\bibfnamefont {L.~J.}\ \bibnamefont
  {Sham}},\ }\bibfield  {title} {\enquote {\bibinfo {title} {Self-consistent
  equations including exchange and correlation effects},}\ }\href {\doibase
  10.1103/PhysRev.140.A1133} {\bibfield  {journal} {\bibinfo  {journal} {Phys.
  Rev.}\ }\textbf {\bibinfo {volume} {140}},\ \bibinfo {pages} {A1133}
  (\bibinfo {year} {1965})}\BibitemShut {NoStop}%
\bibitem [{\citenamefont {Ceperley}\ and\ \citenamefont {Alder}(1980)}]{CA}%
  \BibitemOpen
  \bibfield  {author} {\bibinfo {author} {\bibfnamefont {D.~M.}\ \bibnamefont
  {Ceperley}}\ and\ \bibinfo {author} {\bibfnamefont {B.~J.}\ \bibnamefont
  {Alder}},\ }\bibfield  {title} {\enquote {\bibinfo {title} {Ground state of
  the electron gas by a stochastic method},}\ }\href {\doibase
  10.1103/PhysRevLett.45.566} {\bibfield  {journal} {\bibinfo  {journal} {Phys.
  Rev. Lett.}\ }\textbf {\bibinfo {volume} {45}},\ \bibinfo {pages} {566}
  (\bibinfo {year} {1980})}\BibitemShut {NoStop}%
\bibitem [{\citenamefont {Perdew}\ and\ \citenamefont
  {Schmidt}(2001)}]{jacobsladder}%
  \BibitemOpen
  \bibfield  {author} {\bibinfo {author} {\bibfnamefont {J.~P.}\ \bibnamefont
  {Perdew}}\ and\ \bibinfo {author} {\bibfnamefont {K.}~\bibnamefont
  {Schmidt}},\ }\bibfield  {title} {\enquote {\bibinfo {title} {Jacob’s
  ladder of density functional approximations for the exchange-correlation
  energy},}\ }\href {\doibase 10.1063/1.1390175} {\bibfield  {journal}
  {\bibinfo  {journal} {AIP Conf. Proc.}\ }\textbf {\bibinfo {volume} {577}},\
  \bibinfo {pages} {1} (\bibinfo {year} {2001})}\BibitemShut {NoStop}%
\bibitem [{\citenamefont {Vosko}\ \emph {et~al.}(1980)\citenamefont {Vosko},
  \citenamefont {Wilk},\ and\ \citenamefont {Nusair}}]{VWN}%
  \BibitemOpen
  \bibfield  {author} {\bibinfo {author} {\bibfnamefont {S.~H.}\ \bibnamefont
  {Vosko}}, \bibinfo {author} {\bibfnamefont {L.}~\bibnamefont {Wilk}}, \ and\
  \bibinfo {author} {\bibfnamefont {M.}~\bibnamefont {Nusair}},\ }\bibfield
  {title} {\enquote {\bibinfo {title} {Accurate spin-dependent electron liquid
  correlation energies for local spin density calculations: a critical
  analysis},}\ }\href {\doibase 10.1139/p80-159} {\bibfield  {journal}
  {\bibinfo  {journal} {Can. J. Phys.}\ }\textbf {\bibinfo {volume} {58}},\
  \bibinfo {pages} {1200} (\bibinfo {year} {1980})}\BibitemShut {NoStop}%
\bibitem [{\citenamefont {Cole}\ and\ \citenamefont {Perdew}(1982)}]{CP}%
  \BibitemOpen
  \bibfield  {author} {\bibinfo {author} {\bibfnamefont {L.~A.}\ \bibnamefont
  {Cole}}\ and\ \bibinfo {author} {\bibfnamefont {J.~P.}\ \bibnamefont
  {Perdew}},\ }\bibfield  {title} {\enquote {\bibinfo {title} {Calculated
  electron affinities of the elements},}\ }\href {\doibase
  10.1103/PhysRevA.25.1265} {\bibfield  {journal} {\bibinfo  {journal} {Phys.
  Rev. A}\ }\textbf {\bibinfo {volume} {25}},\ \bibinfo {pages} {1265}
  (\bibinfo {year} {1982})}\BibitemShut {NoStop}%
\bibitem [{\citenamefont {Perdew}\ and\ \citenamefont {Wang}(1992)}]{PW}%
  \BibitemOpen
  \bibfield  {author} {\bibinfo {author} {\bibfnamefont {J.~P.}\ \bibnamefont
  {Perdew}}\ and\ \bibinfo {author} {\bibfnamefont {Y.}~\bibnamefont {Wang}},\
  }\bibfield  {title} {\enquote {\bibinfo {title} {Accurate and simple analytic
  representation of the electron-gas correlation energy},}\ }\href {\doibase
  10.1103/PhysRevB.45.13244} {\bibfield  {journal} {\bibinfo  {journal} {Phys.
  Rev. B}\ }\textbf {\bibinfo {volume} {45}},\ \bibinfo {pages} {13244}
  (\bibinfo {year} {1992})}\BibitemShut {NoStop}%
\bibitem [{\citenamefont {Perdew}\ \emph
  {et~al.}(1996{\natexlab{a}})\citenamefont {Perdew}, \citenamefont {Burke},\
  and\ \citenamefont {Ernzerhof}}]{PBE}%
  \BibitemOpen
  \bibfield  {author} {\bibinfo {author} {\bibfnamefont {J.~P.}\ \bibnamefont
  {Perdew}}, \bibinfo {author} {\bibfnamefont {K.}~\bibnamefont {Burke}}, \
  and\ \bibinfo {author} {\bibfnamefont {M.}~\bibnamefont {Ernzerhof}},\
  }\bibfield  {title} {\enquote {\bibinfo {title} {Generalized gradient
  approximation made simple},}\ }\href {\doibase 10.1103/PhysRevLett.77.3865}
  {\bibfield  {journal} {\bibinfo  {journal} {Phys. Rev. Lett.}\ }\textbf
  {\bibinfo {volume} {77}},\ \bibinfo {pages} {3865} (\bibinfo {year}
  {1996}{\natexlab{a}})}\BibitemShut {NoStop}%
\bibitem [{\citenamefont {Perdew}\ \emph
  {et~al.}(1996{\natexlab{b}})\citenamefont {Perdew}, \citenamefont {Burke},\
  and\ \citenamefont {Wang}}]{PBW96}%
  \BibitemOpen
  \bibfield  {author} {\bibinfo {author} {\bibfnamefont {J.~P.}\ \bibnamefont
  {Perdew}}, \bibinfo {author} {\bibfnamefont {K.}~\bibnamefont {Burke}}, \
  and\ \bibinfo {author} {\bibfnamefont {Y.}~\bibnamefont {Wang}},\ }\bibfield
  {title} {\enquote {\bibinfo {title} {Generalized gradient approximation for
  the exchange-correlation hole of a many-electron system},}\ }\href {\doibase
  10.1103/PhysRevB.54.16533} {\bibfield  {journal} {\bibinfo  {journal} {Phys.
  Rev. B}\ }\textbf {\bibinfo {volume} {54}},\ \bibinfo {pages} {16533}
  (\bibinfo {year} {1996}{\natexlab{b}})}\BibitemShut {NoStop}%
\bibitem [{\citenamefont {Sun}\ \emph {et~al.}(2016)\citenamefont {Sun},
  \citenamefont {Remsing}, \citenamefont {Zhang}, \citenamefont {Sun},
  \citenamefont {Ruzsinszky}, \citenamefont {Peng}, \citenamefont {Yang},
  \citenamefont {Paul}, \citenamefont {Waghmare}, \citenamefont {Wu},
  \citenamefont {Klein},\ and\ \citenamefont {Perdew}}]{Sun2016}%
  \BibitemOpen
  \bibfield  {author} {\bibinfo {author} {\bibfnamefont {J.}~\bibnamefont
  {Sun}}, \bibinfo {author} {\bibfnamefont {R.~C.}\ \bibnamefont {Remsing}},
  \bibinfo {author} {\bibfnamefont {Y.}~\bibnamefont {Zhang}}, \bibinfo
  {author} {\bibfnamefont {Z.}~\bibnamefont {Sun}}, \bibinfo {author}
  {\bibfnamefont {A.}~\bibnamefont {Ruzsinszky}}, \bibinfo {author}
  {\bibfnamefont {H.}~\bibnamefont {Peng}}, \bibinfo {author} {\bibfnamefont
  {Z.}~\bibnamefont {Yang}}, \bibinfo {author} {\bibfnamefont {A.}~\bibnamefont
  {Paul}}, \bibinfo {author} {\bibfnamefont {U.}~\bibnamefont {Waghmare}},
  \bibinfo {author} {\bibfnamefont {X.}~\bibnamefont {Wu}}, \bibinfo {author}
  {\bibfnamefont {M.~L.}\ \bibnamefont {Klein}}, \ and\ \bibinfo {author}
  {\bibfnamefont {J.~P.}\ \bibnamefont {Perdew}},\ }\bibfield  {title}
  {\enquote {\bibinfo {title} {Accurate first-principles structures and
  energies of diversely bonded systems from an efficient density functional},}\
  }\href {\doibase 10.1038/nchem.2535} {\bibfield  {journal} {\bibinfo
  {journal} {Nat. Chem.}\ }\textbf {\bibinfo {volume} {8}},\ \bibinfo {pages}
  {831} (\bibinfo {year} {2016})}\BibitemShut {NoStop}%
\bibitem [{\citenamefont {Becke}(1993)}]{doi:10.1063/1.464304}%
  \BibitemOpen
  \bibfield  {author} {\bibinfo {author} {\bibfnamefont {A.~D.}\ \bibnamefont
  {Becke}},\ }\bibfield  {title} {\enquote {\bibinfo {title} {{A new mixing of
  Hartree–Fock and local density‐functional theories}},}\ }\href {\doibase
  10.1063/1.464304} {\bibfield  {journal} {\bibinfo  {journal} {J. Chem.
  Phys.}\ }\textbf {\bibinfo {volume} {98}},\ \bibinfo {pages} {1372} (\bibinfo
  {year} {1993})}\BibitemShut {NoStop}%
\bibitem [{\citenamefont {Perdew}\ \emph
  {et~al.}(1996{\natexlab{c}})\citenamefont {Perdew}, \citenamefont
  {Ernzerhof},\ and\ \citenamefont {Burke}}]{doi:10.1063/1.472933}%
  \BibitemOpen
  \bibfield  {author} {\bibinfo {author} {\bibfnamefont {J.~P.}\ \bibnamefont
  {Perdew}}, \bibinfo {author} {\bibfnamefont {M.}~\bibnamefont {Ernzerhof}}, \
  and\ \bibinfo {author} {\bibfnamefont {K.}~\bibnamefont {Burke}},\ }\bibfield
   {title} {\enquote {\bibinfo {title} {Rationale for mixing exact exchange
  with density functional approximations},}\ }\href {\doibase 10.1063/1.472933}
  {\bibfield  {journal} {\bibinfo  {journal} {J. Chem. Phys.}\ }\textbf
  {\bibinfo {volume} {105}},\ \bibinfo {pages} {9982} (\bibinfo {year}
  {1996}{\natexlab{c}})}\BibitemShut {NoStop}%
\bibitem [{\citenamefont {Adamo}\ and\ \citenamefont
  {Barone}(1999)}]{doi:10.1063/1.478522}%
  \BibitemOpen
  \bibfield  {author} {\bibinfo {author} {\bibfnamefont {C.}~\bibnamefont
  {Adamo}}\ and\ \bibinfo {author} {\bibfnamefont {V.}~\bibnamefont {Barone}},\
  }\bibfield  {title} {\enquote {\bibinfo {title} {{Toward reliable density
  func\-tional methods without adjustable parameters: The PBE0 model}},}\
  }\href {\doibase 10.1063/1.478522} {\bibfield  {journal} {\bibinfo  {journal}
  {J. Chem. Phys.}\ }\textbf {\bibinfo {volume} {110}},\ \bibinfo {pages}
  {6158} (\bibinfo {year} {1999})}\BibitemShut {NoStop}%
\bibitem [{\citenamefont {Lehtola}\ \emph {et~al.}(2018)\citenamefont
  {Lehtola}, \citenamefont {Steigemann}, \citenamefont {Oliveira},\ and\
  \citenamefont {Marques}}]{libxc}%
  \BibitemOpen
  \bibfield  {author} {\bibinfo {author} {\bibfnamefont {S.}~\bibnamefont
  {Lehtola}}, \bibinfo {author} {\bibfnamefont {C.}~\bibnamefont {Steigemann}},
  \bibinfo {author} {\bibfnamefont {M.~J.~T.}\ \bibnamefont {Oliveira}}, \ and\
  \bibinfo {author} {\bibfnamefont {M.~A.~L.}\ \bibnamefont {Marques}},\
  }\bibfield  {title} {\enquote {\bibinfo {title} {{Recent developments in
  libxc — A comprehensive library of functionals for density functional
  theory}},}\ }\href {\doibase https://doi.org/10.1016/j.softx.2017.11.002}
  {\bibfield  {journal} {\bibinfo  {journal} {SoftwareX}\ }\textbf {\bibinfo
  {volume} {7}},\ \bibinfo {pages} {1} (\bibinfo {year} {2018})}\BibitemShut
  {NoStop}%
\bibitem [{\citenamefont {Becke}(2014)}]{doi:10.1063/1.4869598}%
  \BibitemOpen
  \bibfield  {author} {\bibinfo {author} {\bibfnamefont {A.~D.}\ \bibnamefont
  {Becke}},\ }\bibfield  {title} {\enquote {\bibinfo {title} {Perspective:
  Fifty years of density-functional theory in chemical physics},}\ }\href
  {\doibase 10.1063/1.4869598} {\bibfield  {journal} {\bibinfo  {journal} {J.
  Chem. Phys.}\ }\textbf {\bibinfo {volume} {140}},\ \bibinfo {pages} {18A301}
  (\bibinfo {year} {2014})}\BibitemShut {NoStop}%
\bibitem [{\citenamefont {Cohen}\ \emph
  {et~al.}(2008{\natexlab{a}})\citenamefont {Cohen}, \citenamefont
  {Mori-S{\'a}nchez},\ and\ \citenamefont {Yang}}]{Cohen792}%
  \BibitemOpen
  \bibfield  {author} {\bibinfo {author} {\bibfnamefont {A.~J.}\ \bibnamefont
  {Cohen}}, \bibinfo {author} {\bibfnamefont {P.}~\bibnamefont
  {Mori-S{\'a}nchez}}, \ and\ \bibinfo {author} {\bibfnamefont
  {W.}~\bibnamefont {Yang}},\ }\bibfield  {title} {\enquote {\bibinfo {title}
  {Insights into current limitations of density functional theory},}\ }\href
  {\doibase 10.1126/science.1158722} {\bibfield  {journal} {\bibinfo  {journal}
  {Science}\ }\textbf {\bibinfo {volume} {321}},\ \bibinfo {pages} {792}
  (\bibinfo {year} {2008}{\natexlab{a}})}\BibitemShut {NoStop}%
\bibitem [{\citenamefont {Cohen}\ \emph
  {et~al.}(2008{\natexlab{b}})\citenamefont {Cohen}, \citenamefont
  {Mori-Sánchez},\ and\ \citenamefont {Yang}}]{CMSY}%
  \BibitemOpen
  \bibfield  {author} {\bibinfo {author} {\bibfnamefont {A.~J.}\ \bibnamefont
  {Cohen}}, \bibinfo {author} {\bibfnamefont {P.}~\bibnamefont
  {Mori-Sánchez}}, \ and\ \bibinfo {author} {\bibfnamefont {W.}~\bibnamefont
  {Yang}},\ }\bibfield  {title} {\enquote {\bibinfo {title} {Fractional spins
  and static correlation error in density functional theory},}\ }\href
  {\doibase 10.1063/1.2987202} {\bibfield  {journal} {\bibinfo  {journal} {J.
  Chem. Phys.}\ }\textbf {\bibinfo {volume} {129}},\ \bibinfo {pages} {121104}
  (\bibinfo {year} {2008}{\natexlab{b}})}\BibitemShut {NoStop}%
\bibitem [{\citenamefont {Skone}\ \emph {et~al.}(2014)\citenamefont {Skone},
  \citenamefont {Govoni},\ and\ \citenamefont {Galli}}]{PhysRevB.89.195112}%
  \BibitemOpen
  \bibfield  {author} {\bibinfo {author} {\bibfnamefont {J.~H.}\ \bibnamefont
  {Skone}}, \bibinfo {author} {\bibfnamefont {M.}~\bibnamefont {Govoni}}, \
  and\ \bibinfo {author} {\bibfnamefont {G.}~\bibnamefont {Galli}},\ }\bibfield
   {title} {\enquote {\bibinfo {title} {Self-consistent hybrid functional for
  condensed systems},}\ }\href {\doibase 10.1103/PhysRevB.89.195112} {\bibfield
   {journal} {\bibinfo  {journal} {Phys. Rev. B}\ }\textbf {\bibinfo {volume}
  {89}},\ \bibinfo {pages} {195112} (\bibinfo {year} {2014})}\BibitemShut
  {NoStop}%
\bibitem [{\citenamefont {Benavides-Riveros}\ \emph {et~al.}(2017)\citenamefont
  {Benavides-Riveros}, \citenamefont {Lathiotakis},\ and\ \citenamefont
  {Marques}}]{C7CP01137G}%
  \BibitemOpen
  \bibfield  {author} {\bibinfo {author} {\bibfnamefont {C.~L.}\ \bibnamefont
  {Benavides-Riveros}}, \bibinfo {author} {\bibfnamefont {N.~N.}\ \bibnamefont
  {Lathiotakis}}, \ and\ \bibinfo {author} {\bibfnamefont {M.~A.~L.}\
  \bibnamefont {Marques}},\ }\bibfield  {title} {\enquote {\bibinfo {title}
  {Towards a formal definition of static and dynamic electronic
  correlations},}\ }\href {\doibase 10.1039/C7CP01137G} {\bibfield  {journal}
  {\bibinfo  {journal} {Phys. Chem. Chem. Phys.}\ }\textbf {\bibinfo {volume}
  {19}},\ \bibinfo {pages} {12655} (\bibinfo {year} {2017})}\BibitemShut
  {NoStop}%
\bibitem [{\citenamefont {Liu}\ and\ \citenamefont
  {Tian}(2010)}]{liu2010facial}%
  \BibitemOpen
  \bibfield  {author} {\bibinfo {author} {\bibfnamefont {S.-S.}\ \bibnamefont
  {Liu}}\ and\ \bibinfo {author} {\bibfnamefont {Y.-T.}\ \bibnamefont {Tian}},\
  }\bibfield  {title} {\enquote {\bibinfo {title} {Facial expression
  recognition method based on gabor wavelet features and fractional power
  polynomial kernel {PCA}},}\ }in\ \href {\doibase
  10.1007/978-3-642-13318-3_19} {\emph {\bibinfo {booktitle} {{}Advances in
  Neural Networks}}}\ (\bibinfo  {publisher} {Springer Berlin Heidelberg},\
  \bibinfo {year} {2010})\ pp.\ \bibinfo {pages} {144--151}\BibitemShut
  {NoStop}%
\bibitem [{\citenamefont {Waibel}\ and\ \citenamefont
  {Lee}(1990)}]{waibel1990readings}%
  \BibitemOpen
  \bibinfo {editor} {\bibfnamefont {A.}~\bibnamefont {Waibel}}\ and\ \bibinfo
  {editor} {\bibfnamefont {K.-F.}\ \bibnamefont {Lee}},\ eds.,\ \href
  {https://www.ebook.de/de/product/21119798/readings_in_speech_recognition.html}
  {\emph {\bibinfo {title} {Readings in Speech Recognition}}}\ (\bibinfo
  {publisher} {Morgan Kaufmann},\ \bibinfo {year} {1990})\BibitemShut {NoStop}%
\bibitem [{\citenamefont {Butler}\ \emph {et~al.}(2018)\citenamefont {Butler},
  \citenamefont {Davies}, \citenamefont {Cartwright}, \citenamefont {Isayev},\
  and\ \citenamefont {Walsh}}]{butler2019}%
  \BibitemOpen
  \bibfield  {author} {\bibinfo {author} {\bibfnamefont {K.~T.}\ \bibnamefont
  {Butler}}, \bibinfo {author} {\bibfnamefont {D.~W.}\ \bibnamefont {Davies}},
  \bibinfo {author} {\bibfnamefont {H.}~\bibnamefont {Cartwright}}, \bibinfo
  {author} {\bibfnamefont {O.}~\bibnamefont {Isayev}}, \ and\ \bibinfo {author}
  {\bibfnamefont {A.}~\bibnamefont {Walsh}},\ }\bibfield  {title} {\enquote
  {\bibinfo {title} {Machine learning for molecular and materials science},}\
  }\href {\doibase 10.1038/s41586-018-0337-2} {\bibfield  {journal} {\bibinfo
  {journal} {Nature}\ }\textbf {\bibinfo {volume} {559}},\ \bibinfo {pages}
  {547} (\bibinfo {year} {2018})}\BibitemShut {NoStop}%
\bibitem [{\citenamefont {Schmidt}\ \emph {et~al.}(2019)\citenamefont
  {Schmidt}, \citenamefont {Marques}, \citenamefont {Botti},\ and\
  \citenamefont {Marques}}]{ourreview}%
  \BibitemOpen
  \bibfield  {author} {\bibinfo {author} {\bibfnamefont {J.}~\bibnamefont
  {Schmidt}}, \bibinfo {author} {\bibfnamefont {M.~R.~G.}\ \bibnamefont
  {Marques}}, \bibinfo {author} {\bibfnamefont {S.}~\bibnamefont {Botti}}, \
  and\ \bibinfo {author} {\bibfnamefont {M.~A.~L.}\ \bibnamefont {Marques}},\
  }\bibfield  {title} {\enquote {\bibinfo {title} {Recent advances and
  applications of machine learning in solid-state materials science},}\ }\href
  {https://doi.org/10.1038/s41524-019-0221-0} {\bibfield  {journal} {\bibinfo
  {journal} {Npj Comput. Mat.}\ }\textbf {\bibinfo {volume} {5}} (\bibinfo
  {year} {2019})}\BibitemShut {NoStop}%
\bibitem [{\citenamefont {Ghiringhelli}\ \emph {et~al.}(2015)\citenamefont
  {Ghiringhelli}, \citenamefont {Vybiral}, \citenamefont {Levchenko},
  \citenamefont {Draxl},\ and\ \citenamefont
  {Scheffler}}]{PhysRevLett.114.105503}%
  \BibitemOpen
  \bibfield  {author} {\bibinfo {author} {\bibfnamefont {L.~M.}\ \bibnamefont
  {Ghiringhelli}}, \bibinfo {author} {\bibfnamefont {J.}~\bibnamefont
  {Vybiral}}, \bibinfo {author} {\bibfnamefont {S.~V.}\ \bibnamefont
  {Levchenko}}, \bibinfo {author} {\bibfnamefont {C.}~\bibnamefont {Draxl}}, \
  and\ \bibinfo {author} {\bibfnamefont {M.}~\bibnamefont {Scheffler}},\
  }\bibfield  {title} {\enquote {\bibinfo {title} {Big data of materials
  science: Critical role of the descriptor},}\ }\href {\doibase
  10.1103/PhysRevLett.114.105503} {\bibfield  {journal} {\bibinfo  {journal}
  {Phys. Rev. Lett.}\ }\textbf {\bibinfo {volume} {114}},\ \bibinfo {pages}
  {105503} (\bibinfo {year} {2015})}\BibitemShut {NoStop}%
\bibitem [{\citenamefont {Tozer}\ \emph {et~al.}(1996)\citenamefont {Tozer},
  \citenamefont {Ingamells},\ and\ \citenamefont {Handy}}]{tozer1996exchange}%
  \BibitemOpen
  \bibfield  {author} {\bibinfo {author} {\bibfnamefont {D.~J.}\ \bibnamefont
  {Tozer}}, \bibinfo {author} {\bibfnamefont {V.~E.}\ \bibnamefont
  {Ingamells}}, \ and\ \bibinfo {author} {\bibfnamefont {N.~C.}\ \bibnamefont
  {Handy}},\ }\bibfield  {title} {\enquote {\bibinfo {title}
  {Exchange-correlation potentials},}\ }\href {\doibase 10.1063/1.472753}
  {\bibfield  {journal} {\bibinfo  {journal} {J. Chem. Phys.}\ }\textbf
  {\bibinfo {volume} {105}},\ \bibinfo {pages} {9200} (\bibinfo {year}
  {1996})}\BibitemShut {NoStop}%
\bibitem [{\citenamefont {Snyder}\ \emph {et~al.}(2012)\citenamefont {Snyder},
  \citenamefont {Rupp}, \citenamefont {Hansen}, \citenamefont {Müller},\ and\
  \citenamefont {Burke}}]{snyder2012finding}%
  \BibitemOpen
  \bibfield  {author} {\bibinfo {author} {\bibfnamefont {J.~C.}\ \bibnamefont
  {Snyder}}, \bibinfo {author} {\bibfnamefont {M.}~\bibnamefont {Rupp}},
  \bibinfo {author} {\bibfnamefont {K.}~\bibnamefont {Hansen}}, \bibinfo
  {author} {\bibfnamefont {K-R.}\ \bibnamefont {Müller}}, \ and\ \bibinfo
  {author} {\bibfnamefont {K.}~\bibnamefont {Burke}},\ }\bibfield  {title}
  {\enquote {\bibinfo {title} {Finding density functionals with machine
  learning},}\ }\href {\doibase 10.1103/physrevlett.108.253002} {\bibfield
  {journal} {\bibinfo  {journal} {Phys. Rev. Lett.}\ }\textbf {\bibinfo
  {volume} {108}},\ \bibinfo {pages} {253002} (\bibinfo {year}
  {2012})}\BibitemShut {NoStop}%
\bibitem [{\citenamefont {Snyder}\ \emph {et~al.}(2013)\citenamefont {Snyder},
  \citenamefont {Rupp}, \citenamefont {Hansen}, \citenamefont {Blooston},
  \citenamefont {Müller},\ and\ \citenamefont {Burke}}]{snyder2013orbital}%
  \BibitemOpen
  \bibfield  {author} {\bibinfo {author} {\bibfnamefont {J.~C.}\ \bibnamefont
  {Snyder}}, \bibinfo {author} {\bibfnamefont {M.}~\bibnamefont {Rupp}},
  \bibinfo {author} {\bibfnamefont {K.}~\bibnamefont {Hansen}}, \bibinfo
  {author} {\bibfnamefont {L.}~\bibnamefont {Blooston}}, \bibinfo {author}
  {\bibfnamefont {K.-R.}\ \bibnamefont {Müller}}, \ and\ \bibinfo {author}
  {\bibfnamefont {K.}~\bibnamefont {Burke}},\ }\bibfield  {title} {\enquote
  {\bibinfo {title} {Orbital-free bond breaking via machine learning},}\ }\href
  {\doibase 10.1063/1.4834075} {\bibfield  {journal} {\bibinfo  {journal} {J.
  Chem. Phys.}\ }\textbf {\bibinfo {volume} {139}},\ \bibinfo {pages} {224104}
  (\bibinfo {year} {2013})}\BibitemShut {NoStop}%
\bibitem [{\citenamefont {Yao}\ and\ \citenamefont
  {Parkhill}(2016)}]{doi:10.1021/acs.jctc.5b01011}%
  \BibitemOpen
  \bibfield  {author} {\bibinfo {author} {\bibfnamefont {K.}~\bibnamefont
  {Yao}}\ and\ \bibinfo {author} {\bibfnamefont {J.}~\bibnamefont {Parkhill}},\
  }\bibfield  {title} {\enquote {\bibinfo {title} {{Kinetic Energy of
  Hydrocarbons as a Function of Electron Density and Convolutional Neural
  Networks}},}\ }\href {\doibase 10.1021/acs.jctc.5b01011} {\bibfield
  {journal} {\bibinfo  {journal} {J. Chem. Theory Comput.}\ }\textbf {\bibinfo
  {volume} {12}},\ \bibinfo {pages} {1139} (\bibinfo {year}
  {2016})}\BibitemShut {NoStop}%
\bibitem [{\citenamefont {Brockherde}\ \emph {et~al.}(2017)\citenamefont
  {Brockherde}, \citenamefont {Vogt}, \citenamefont {Li}, \citenamefont
  {Tuckerman}, \citenamefont {Burke},\ and\ \citenamefont
  {Müller}}]{brockherde2017bypassing}%
  \BibitemOpen
  \bibfield  {author} {\bibinfo {author} {\bibfnamefont {F.}~\bibnamefont
  {Brockherde}}, \bibinfo {author} {\bibfnamefont {L.}~\bibnamefont {Vogt}},
  \bibinfo {author} {\bibfnamefont {L.}~\bibnamefont {Li}}, \bibinfo {author}
  {\bibfnamefont {M.~E.}\ \bibnamefont {Tuckerman}}, \bibinfo {author}
  {\bibfnamefont {K.}~\bibnamefont {Burke}}, \ and\ \bibinfo {author}
  {\bibfnamefont {K.-R.}\ \bibnamefont {Müller}},\ }\bibfield  {title}
  {\enquote {\bibinfo {title} {Bypassing the {Kohn-Sham} equations with machine
  learning},}\ }\href {\doibase 10.1038/s41467-017-00839-3} {\bibfield
  {journal} {\bibinfo  {journal} {Nat. Commun.}\ }\textbf {\bibinfo {volume}
  {8}},\ \bibinfo {pages} {872} (\bibinfo {year} {2017})}\BibitemShut {NoStop}%
\bibitem [{\citenamefont {Liu}\ \emph {et~al.}(2017)\citenamefont {Liu},
  \citenamefont {Wang}, \citenamefont {Du}, \citenamefont {Hu}, \citenamefont
  {Zheng},\ and\ \citenamefont {Chen}}]{liu2017improving}%
  \BibitemOpen
  \bibfield  {author} {\bibinfo {author} {\bibfnamefont {Q.}~\bibnamefont
  {Liu}}, \bibinfo {author} {\bibfnamefont {J.~C.}\ \bibnamefont {Wang}},
  \bibinfo {author} {\bibfnamefont {P.~L.}\ \bibnamefont {Du}}, \bibinfo
  {author} {\bibfnamefont {L.~H.}\ \bibnamefont {Hu}}, \bibinfo {author}
  {\bibfnamefont {X.}~\bibnamefont {Zheng}}, \ and\ \bibinfo {author}
  {\bibfnamefont {G.}~\bibnamefont {Chen}},\ }\bibfield  {title} {\enquote
  {\bibinfo {title} {Improving the performance of long-range-corrected
  exchange-correlation functional with an embedded neural network},}\ }\href
  {\doibase 10.1021/acs.jpca.7b07045} {\bibfield  {journal} {\bibinfo
  {journal} {J. Phys. Chem. A}\ }\textbf {\bibinfo {volume} {121}},\ \bibinfo
  {pages} {7273} (\bibinfo {year} {2017})}\BibitemShut {NoStop}%
\bibitem [{\citenamefont {Nagal}\ \emph {et~al.}(2018)\citenamefont {Nagal},
  \citenamefont {Akashi}, \citenamefont {Sasaki},\ and\ \citenamefont
  {Tsuneyuki}}]{nagai2018neural}%
  \BibitemOpen
  \bibfield  {author} {\bibinfo {author} {\bibfnamefont {R.}~\bibnamefont
  {Nagal}}, \bibinfo {author} {\bibfnamefont {R.}~\bibnamefont {Akashi}},
  \bibinfo {author} {\bibfnamefont {S.}~\bibnamefont {Sasaki}}, \ and\ \bibinfo
  {author} {\bibfnamefont {S.}~\bibnamefont {Tsuneyuki}},\ }\bibfield  {title}
  {\enquote {\bibinfo {title} {Neural-network {Kohn-Sham} exchange-correlation
  potential and its out-of-training transferability},}\ }\href {\doibase
  10.1063/1.5029279} {\bibfield  {journal} {\bibinfo  {journal} {J. Chem.
  Phys.}\ }\textbf {\bibinfo {volume} {148}},\ \bibinfo {pages} {241737}
  (\bibinfo {year} {2018})}\BibitemShut {NoStop}%
\bibitem [{\citenamefont {Lundgaard}\ \emph {et~al.}(2016)\citenamefont
  {Lundgaard}, \citenamefont {Wellendorff}, \citenamefont {Voss}, \citenamefont
  {Jacobsen},\ and\ \citenamefont {Bligaard}}]{PhysRevB.93.235162}%
  \BibitemOpen
  \bibfield  {author} {\bibinfo {author} {\bibfnamefont {K.~T.}\ \bibnamefont
  {Lundgaard}}, \bibinfo {author} {\bibfnamefont {J.}~\bibnamefont
  {Wellendorff}}, \bibinfo {author} {\bibfnamefont {J.}~\bibnamefont {Voss}},
  \bibinfo {author} {\bibfnamefont {K.~W.}\ \bibnamefont {Jacobsen}}, \ and\
  \bibinfo {author} {\bibfnamefont {T.}~\bibnamefont {Bligaard}},\ }\bibfield
  {title} {\enquote {\bibinfo {title} {{mBEEF-vdW: Robust fitting of error
  estimation density functionals}},}\ }\href {\doibase
  10.1103/PhysRevB.93.235162} {\bibfield  {journal} {\bibinfo  {journal} {Phys.
  Rev. B}\ }\textbf {\bibinfo {volume} {93}},\ \bibinfo {pages} {235162}
  (\bibinfo {year} {2016})}\BibitemShut {NoStop}%
\bibitem [{\citenamefont {Wellendorff}\ \emph {et~al.}(2012)\citenamefont
  {Wellendorff}, \citenamefont {Lundgaard}, \citenamefont {M\o{}gelh\o{}j},
  \citenamefont {Petzold}, \citenamefont {Landis}, \citenamefont {N\o{}rskov},
  \citenamefont {Bligaard},\ and\ \citenamefont
  {Jacobsen}}]{PhysRevB.85.235149}%
  \BibitemOpen
  \bibfield  {author} {\bibinfo {author} {\bibfnamefont {J.}~\bibnamefont
  {Wellendorff}}, \bibinfo {author} {\bibfnamefont {K.~T.}\ \bibnamefont
  {Lundgaard}}, \bibinfo {author} {\bibfnamefont {A.}~\bibnamefont
  {M\o{}gelh\o{}j}}, \bibinfo {author} {\bibfnamefont {V.}~\bibnamefont
  {Petzold}}, \bibinfo {author} {\bibfnamefont {D.~D.}\ \bibnamefont {Landis}},
  \bibinfo {author} {\bibfnamefont {J.~K.}\ \bibnamefont {N\o{}rskov}},
  \bibinfo {author} {\bibfnamefont {T.}~\bibnamefont {Bligaard}}, \ and\
  \bibinfo {author} {\bibfnamefont {K.~W.}\ \bibnamefont {Jacobsen}},\
  }\bibfield  {title} {\enquote {\bibinfo {title} {Density functionals for
  surface science: Exchange-correlation model development with bayesian error
  estimation},}\ }\href {\doibase 10.1103/PhysRevB.85.235149} {\bibfield
  {journal} {\bibinfo  {journal} {Phys. Rev. B}\ }\textbf {\bibinfo {volume}
  {85}},\ \bibinfo {pages} {235149} (\bibinfo {year} {2012})}\BibitemShut
  {NoStop}%
\bibitem [{\citenamefont {Li}\ \emph {et~al.}(2016)\citenamefont {Li},
  \citenamefont {Baker}, \citenamefont {White},\ and\ \citenamefont
  {Burke}}]{PhysRevB.94.245129}%
  \BibitemOpen
  \bibfield  {author} {\bibinfo {author} {\bibfnamefont {L.}~\bibnamefont
  {Li}}, \bibinfo {author} {\bibfnamefont {T.~E.}\ \bibnamefont {Baker}},
  \bibinfo {author} {\bibfnamefont {S.~R.}\ \bibnamefont {White}}, \ and\
  \bibinfo {author} {\bibfnamefont {K.}~\bibnamefont {Burke}},\ }\bibfield
  {title} {\enquote {\bibinfo {title} {Pure den\-si\-ty functional for strong
  correlation and the thermodynamic limit from machine learning},}\ }\href
  {\doibase 10.1103/PhysRevB.94.245129} {\bibfield  {journal} {\bibinfo
  {journal} {Phys. Rev. B}\ }\textbf {\bibinfo {volume} {94}},\ \bibinfo
  {pages} {245129} (\bibinfo {year} {2016})}\BibitemShut {NoStop}%
\bibitem [{\citenamefont {Krieger}\ \emph {et~al.}(1990)\citenamefont
  {Krieger}, \citenamefont {Li},\ and\ \citenamefont {Iafrate}}]{KLI}%
  \BibitemOpen
  \bibfield  {author} {\bibinfo {author} {\bibfnamefont {J.~B.}\ \bibnamefont
  {Krieger}}, \bibinfo {author} {\bibfnamefont {Y.}~\bibnamefont {Li}}, \ and\
  \bibinfo {author} {\bibfnamefont {G.~J.}\ \bibnamefont {Iafrate}},\
  }\bibfield  {title} {\enquote {\bibinfo {title} {{Derivation and application
  of an accurate Kohn-Sham potential with integer discontinuity}},}\ }\href
  {http://www.sciencedirect.com/science/article/pii/037596019090975T}
  {\bibfield  {journal} {\bibinfo  {journal} {Phys. Lett. A}\ }\textbf
  {\bibinfo {volume} {146}},\ \bibinfo {pages} {256} (\bibinfo {year}
  {1990})}\BibitemShut {NoStop}%
\bibitem [{\citenamefont {Sharp}\ and\ \citenamefont
  {Horton}(1953)}]{Sharp_1953}%
  \BibitemOpen
  \bibfield  {author} {\bibinfo {author} {\bibfnamefont {R.~T.}\ \bibnamefont
  {Sharp}}\ and\ \bibinfo {author} {\bibfnamefont {G.~K.}\ \bibnamefont
  {Horton}},\ }\bibfield  {title} {\enquote {\bibinfo {title} {A variational
  approach to the unipotential many-electron problem},}\ }\href {\doibase
  10.1103/PhysRev.90.317} {\bibfield  {journal} {\bibinfo  {journal} {Phys.
  Rev.}\ }\textbf {\bibinfo {volume} {90}},\ \bibinfo {pages} {317} (\bibinfo
  {year} {1953})}\BibitemShut {NoStop}%
\bibitem [{\citenamefont {Talman}\ and\ \citenamefont
  {Shadwick}(1976)}]{Talman_1976}%
  \BibitemOpen
  \bibfield  {author} {\bibinfo {author} {\bibfnamefont {J.~D.}\ \bibnamefont
  {Talman}}\ and\ \bibinfo {author} {\bibfnamefont {W.~F.}\ \bibnamefont
  {Shadwick}},\ }\bibfield  {title} {\enquote {\bibinfo {title} {Optimized
  effective atomic central potential},}\ }\href {\doibase
  10.1103/PhysRevA.14.36} {\bibfield  {journal} {\bibinfo  {journal} {Phys.
  Rev. A}\ }\textbf {\bibinfo {volume} {14}},\ \bibinfo {pages} {36} (\bibinfo
  {year} {1976})}\BibitemShut {NoStop}%
\bibitem [{\citenamefont {van Leeuwen}\ and\ \citenamefont
  {Baerends}(1994)}]{LB94}%
  \BibitemOpen
  \bibfield  {author} {\bibinfo {author} {\bibfnamefont {R.}~\bibnamefont {van
  Leeuwen}}\ and\ \bibinfo {author} {\bibfnamefont {E.~J.}\ \bibnamefont
  {Baerends}},\ }\bibfield  {title} {\enquote {\bibinfo {title}
  {Exchange-co\-rre\-lation potential with correct asymptotic behavior},}\
  }\href {\doibase 10.1103/PhysRevA.49.2421} {\bibfield  {journal} {\bibinfo
  {journal} {Phys. Rev. A}\ }\textbf {\bibinfo {volume} {49}},\ \bibinfo
  {pages} {2421} (\bibinfo {year} {1994})}\BibitemShut {NoStop}%
\bibitem [{\citenamefont {Becke}\ and\ \citenamefont {Johnson}(2006)}]{BJ}%
  \BibitemOpen
  \bibfield  {author} {\bibinfo {author} {\bibfnamefont {A.~D.}\ \bibnamefont
  {Becke}}\ and\ \bibinfo {author} {\bibfnamefont {E.~R.}\ \bibnamefont
  {Johnson}},\ }\bibfield  {title} {\enquote {\bibinfo {title} {A simple
  effective potential for exchange},}\ }\href
  {https://doi.org/10.1063/1.2213970} {\bibfield  {journal} {\bibinfo
  {journal} {J. Chem. Phys.}\ }\textbf {\bibinfo {volume} {124}},\ \bibinfo
  {pages} {221101} (\bibinfo {year} {2006})}\BibitemShut {NoStop}%
\bibitem [{\citenamefont {Tran}\ and\ \citenamefont {Blaha}(2009)}]{TB09}%
  \BibitemOpen
  \bibfield  {author} {\bibinfo {author} {\bibfnamefont {F.}~\bibnamefont
  {Tran}}\ and\ \bibinfo {author} {\bibfnamefont {P.}~\bibnamefont {Blaha}},\
  }\bibfield  {title} {\enquote {\bibinfo {title} {Accurate {{Band Gaps}} of
  {{Semiconductors}} and {{Insulators}} with a {{Semilocal
  Exchange}}-{{Correlation Potential}}},}\ }\href
  {https://link.aps.org/doi/10.1103/PhysRevLett.102.226401} {\bibfield
  {journal} {\bibinfo  {journal} {Phys. Rev. Lett.}\ }\textbf {\bibinfo
  {volume} {102}},\ \bibinfo {pages} {226401} (\bibinfo {year}
  {2009})}\BibitemShut {NoStop}%
\bibitem [{\citenamefont {Gaiduk}\ and\ \citenamefont
  {Staroverov}(2009)}]{gaiduk}%
  \BibitemOpen
  \bibfield  {author} {\bibinfo {author} {\bibfnamefont {A.~P.}\ \bibnamefont
  {Gaiduk}}\ and\ \bibinfo {author} {\bibfnamefont {V.~N.}\ \bibnamefont
  {Staroverov}},\ }\bibfield  {title} {\enquote {\bibinfo {title} {{How to tell
  when a model Kohn–Sham potential is not a functional derivative}},}\ }\href
  {\doibase 10.1063/1.3176515} {\bibfield  {journal} {\bibinfo  {journal} {J.
  Chem. Phys.}\ }\textbf {\bibinfo {volume} {131}},\ \bibinfo {pages} {044107}
  (\bibinfo {year} {2009})}\BibitemShut {NoStop}%
\bibitem [{\citenamefont {Borlido}\ \emph {et~al.}(2019)\citenamefont
  {Borlido}, \citenamefont {Aull}, \citenamefont {Huran}, \citenamefont {Tran},
  \citenamefont {Marques},\ and\ \citenamefont {Botti}}]{usbenchmark}%
  \BibitemOpen
  \bibfield  {author} {\bibinfo {author} {\bibfnamefont {P.}~\bibnamefont
  {Borlido}}, \bibinfo {author} {\bibfnamefont {T.}~\bibnamefont {Aull}},
  \bibinfo {author} {\bibfnamefont {A.~W.}\ \bibnamefont {Huran}}, \bibinfo
  {author} {\bibfnamefont {F.}~\bibnamefont {Tran}}, \bibinfo {author}
  {\bibfnamefont {M.~A.~L.}\ \bibnamefont {Marques}}, \ and\ \bibinfo {author}
  {\bibfnamefont {S.}~\bibnamefont {Botti}},\ }\bibfield  {title} {\enquote
  {\bibinfo {title} {{Large-Scale Benchmark of Ex\-chan\-ge–Correlation
  Functionals for the Determination of Electronic Band Gaps of Solids}},}\
  }\href {\doibase 10.1021/acs.jctc.9b00322} {\bibfield  {journal} {\bibinfo
  {journal} {J. Chem. Theory Comput.}\ }\textbf {\bibinfo {volume} {15}},\
  \bibinfo {pages} {5069} (\bibinfo {year} {2019})}\BibitemShut {NoStop}%
\bibitem [{\citenamefont {Paszke}\ \emph {et~al.}(2017)\citenamefont {Paszke},
  \citenamefont {Gross}, \citenamefont {Chintala}, \citenamefont {Chanan},
  \citenamefont {Yang}, \citenamefont {DeVito}, \citenamefont {Lin},
  \citenamefont {Desmaison}, \citenamefont {Antiga},\ and\ \citenamefont
  {Lerer}}]{paszke2017automatic}%
  \BibitemOpen
  \bibfield  {author} {\bibinfo {author} {\bibfnamefont {A.}~\bibnamefont
  {Paszke}}, \bibinfo {author} {\bibfnamefont {S.}~\bibnamefont {Gross}},
  \bibinfo {author} {\bibfnamefont {S.}~\bibnamefont {Chintala}}, \bibinfo
  {author} {\bibfnamefont {G.}~\bibnamefont {Chanan}}, \bibinfo {author}
  {\bibfnamefont {E.}~\bibnamefont {Yang}}, \bibinfo {author} {\bibfnamefont
  {Z.}~\bibnamefont {DeVito}}, \bibinfo {author} {\bibfnamefont
  {Z.}~\bibnamefont {Lin}}, \bibinfo {author} {\bibfnamefont {A.}~\bibnamefont
  {Desmaison}}, \bibinfo {author} {\bibfnamefont {L.}~\bibnamefont {Antiga}}, \
  and\ \bibinfo {author} {\bibfnamefont {A.}~\bibnamefont {Lerer}},\ }\bibfield
   {title} {\enquote {\bibinfo {title} {Automatic differentiation in
  pytorch},}\ }in\ \href {https://openreview.net/pdf?id=BJJsrmfCZ} {\emph
  {\bibinfo {booktitle} {NIPS 2017 Autodiff Workshop: The Future of
  Gradient-based Machine Learning Software and Techniques}}}\ (\bibinfo {year}
  {2017})\BibitemShut {NoStop}%
\bibitem [{\citenamefont {Abadi}\ \emph {et~al.}(2015)\citenamefont {Abadi},
  \citenamefont {Agarwal}, \citenamefont {Barham}, \citenamefont {Brevdo},
  \citenamefont {Chen}, \citenamefont {Citro}, \citenamefont {Corrado},
  \citenamefont {Davis}, \citenamefont {Dean}, \citenamefont {Devin},
  \citenamefont {Ghemawat}, \citenamefont {Goodfellow}, \citenamefont {Harp},
  \citenamefont {Irving}, \citenamefont {Isard}, \citenamefont {Jia},
  \citenamefont {Jozefowicz}, \citenamefont {Kaiser}, \citenamefont {Kudlur},
  \citenamefont {Levenberg}, \citenamefont {Man\'{e}}, \citenamefont {Monga},
  \citenamefont {Moore}, \citenamefont {Murray}, \citenamefont {Olah},
  \citenamefont {Schuster}, \citenamefont {Shlens}, \citenamefont {Steiner},
  \citenamefont {Sutskever}, \citenamefont {Talwar}, \citenamefont {Tucker},
  \citenamefont {Vanhoucke}, \citenamefont {Vasudevan}, \citenamefont
  {Vi\'{e}gas}, \citenamefont {Vinyals}, \citenamefont {Warden}, \citenamefont
  {Wattenberg}, \citenamefont {Wicke}, \citenamefont {Yu},\ and\ \citenamefont
  {Zheng}}]{tensorflow2015-whitepaper}%
  \BibitemOpen
  \bibfield  {author} {\bibinfo {author} {\bibfnamefont {M.}~\bibnamefont
  {Abadi}}, \bibinfo {author} {\bibfnamefont {A.}~\bibnamefont {Agarwal}},
  \bibinfo {author} {\bibfnamefont {P.}~\bibnamefont {Barham}}, \bibinfo
  {author} {\bibfnamefont {E.}~\bibnamefont {Brevdo}}, \bibinfo {author}
  {\bibfnamefont {Z.}~\bibnamefont {Chen}}, \bibinfo {author} {\bibfnamefont
  {C.}~\bibnamefont {Citro}}, \bibinfo {author} {\bibfnamefont {G.~S.}\
  \bibnamefont {Corrado}}, \bibinfo {author} {\bibfnamefont {A.}~\bibnamefont
  {Davis}}, \bibinfo {author} {\bibfnamefont {J.}~\bibnamefont {Dean}},
  \bibinfo {author} {\bibfnamefont {M.}~\bibnamefont {Devin}}, \bibinfo
  {author} {\bibfnamefont {S.}~\bibnamefont {Ghemawat}}, \bibinfo {author}
  {\bibfnamefont {I.}~\bibnamefont {Goodfellow}}, \bibinfo {author}
  {\bibfnamefont {A.}~\bibnamefont {Harp}}, \bibinfo {author} {\bibfnamefont
  {G.}~\bibnamefont {Irving}}, \bibinfo {author} {\bibfnamefont
  {M.}~\bibnamefont {Isard}}, \bibinfo {author} {\bibfnamefont
  {Y.}~\bibnamefont {Jia}}, \bibinfo {author} {\bibfnamefont {R.}~\bibnamefont
  {Jozefowicz}}, \bibinfo {author} {\bibfnamefont {L.}~\bibnamefont {Kaiser}},
  \bibinfo {author} {\bibfnamefont {M.}~\bibnamefont {Kudlur}}, \bibinfo
  {author} {\bibfnamefont {J.}~\bibnamefont {Levenberg}}, \bibinfo {author}
  {\bibfnamefont {D.}~\bibnamefont {Man\'{e}}}, \bibinfo {author}
  {\bibfnamefont {R.}~\bibnamefont {Monga}}, \bibinfo {author} {\bibfnamefont
  {S.}~\bibnamefont {Moore}}, \bibinfo {author} {\bibfnamefont
  {D.}~\bibnamefont {Murray}}, \bibinfo {author} {\bibfnamefont
  {C.}~\bibnamefont {Olah}}, \bibinfo {author} {\bibfnamefont {M.}~\bibnamefont
  {Schuster}}, \bibinfo {author} {\bibfnamefont {J.}~\bibnamefont {Shlens}},
  \bibinfo {author} {\bibfnamefont {B.}~\bibnamefont {Steiner}}, \bibinfo
  {author} {\bibfnamefont {I.}~\bibnamefont {Sutskever}}, \bibinfo {author}
  {\bibfnamefont {K.}~\bibnamefont {Talwar}}, \bibinfo {author} {\bibfnamefont
  {P.}~\bibnamefont {Tucker}}, \bibinfo {author} {\bibfnamefont
  {V.}~\bibnamefont {Vanhoucke}}, \bibinfo {author} {\bibfnamefont
  {V.}~\bibnamefont {Vasudevan}}, \bibinfo {author} {\bibfnamefont
  {F.}~\bibnamefont {Vi\'{e}gas}}, \bibinfo {author} {\bibfnamefont
  {O.}~\bibnamefont {Vinyals}}, \bibinfo {author} {\bibfnamefont
  {P.}~\bibnamefont {Warden}}, \bibinfo {author} {\bibfnamefont
  {M.}~\bibnamefont {Wattenberg}}, \bibinfo {author} {\bibfnamefont
  {M.}~\bibnamefont {Wicke}}, \bibinfo {author} {\bibfnamefont
  {Y.}~\bibnamefont {Yu}}, \ and\ \bibinfo {author} {\bibfnamefont
  {X.}~\bibnamefont {Zheng}},\ }\href@noop {} {\enquote {\bibinfo {title}
  {{TensorFlow}: Large-scale machine learning on heterogeneous systems},}\
  }\bibinfo {howpublished} {\url{https://tensorflow.org/}} (\bibinfo {year}
  {2015})\BibitemShut {NoStop}%
\bibitem [{\citenamefont {Nagai}\ \emph {et~al.}(2019)\citenamefont {Nagai},
  \citenamefont {Akashi},\ and\ \citenamefont {Sugino}}]{nagai2019completing}%
  \BibitemOpen
  \bibfield  {author} {\bibinfo {author} {\bibfnamefont {R.}~\bibnamefont
  {Nagai}}, \bibinfo {author} {\bibfnamefont {R.}~\bibnamefont {Akashi}}, \
  and\ \bibinfo {author} {\bibfnamefont {O.}~\bibnamefont {Sugino}},\
  }\bibfield  {title} {\enquote {\bibinfo {title} {Completing density
  functional theory by machine-learning hidden messages from molecules},}\
  }\href {https://arxiv.org/abs/1903.00238} {\bibfield  {journal} {\bibinfo
  {journal} {arXiv:1903.00238}\ } (\bibinfo {year} {2019})}\BibitemShut
  {NoStop}%
\bibitem [{\citenamefont {Wagner}\ \emph {et~al.}(2012)\citenamefont {Wagner},
  \citenamefont {Stoudenmire}, \citenamefont {Burke},\ and\ \citenamefont
  {White}}]{C2CP24118H}%
  \BibitemOpen
  \bibfield  {author} {\bibinfo {author} {\bibfnamefont {L.~O.}\ \bibnamefont
  {Wagner}}, \bibinfo {author} {\bibfnamefont {E.~M.}\ \bibnamefont
  {Stoudenmire}}, \bibinfo {author} {\bibfnamefont {K.}~\bibnamefont {Burke}},
  \ and\ \bibinfo {author} {\bibfnamefont {S.~R.}\ \bibnamefont {White}},\
  }\bibfield  {title} {\enquote {\bibinfo {title} {Reference electronic
  structure calculations in one dimension},}\ }\href {\doibase
  10.1039/C2CP24118H} {\bibfield  {journal} {\bibinfo  {journal} {Phys. Chem.
  Chem. Phys.}\ }\textbf {\bibinfo {volume} {14}},\ \bibinfo {pages} {8581}
  (\bibinfo {year} {2012})}\BibitemShut {NoStop}%
\bibitem [{\citenamefont {Andrade}\ \emph {et~al.}(2015)\citenamefont
  {Andrade}, \citenamefont {Strubbe}, \citenamefont {De~Giovannini},
  \citenamefont {Larsen}, \citenamefont {Oliveira}, \citenamefont
  {Alberdi-Rodriguez}, \citenamefont {Varas}, \citenamefont {Theophilou},
  \citenamefont {Helbig}, \citenamefont {Verstraete}, \citenamefont {Stella},
  \citenamefont {Nogueira}, \citenamefont {Aspuru-Guzik}, \citenamefont
  {Castro}, \citenamefont {Marques},\ and\ \citenamefont {Rubio}}]{C5CP00351B}%
  \BibitemOpen
  \bibfield  {author} {\bibinfo {author} {\bibfnamefont {X.}~\bibnamefont
  {Andrade}}, \bibinfo {author} {\bibfnamefont {D.}~\bibnamefont {Strubbe}},
  \bibinfo {author} {\bibfnamefont {U.}~\bibnamefont {De~Giovannini}}, \bibinfo
  {author} {\bibfnamefont {A.~H.}\ \bibnamefont {Larsen}}, \bibinfo {author}
  {\bibfnamefont {M.~J.~T.}\ \bibnamefont {Oliveira}}, \bibinfo {author}
  {\bibfnamefont {J.}~\bibnamefont {Alberdi-Rodriguez}}, \bibinfo {author}
  {\bibfnamefont {A.}~\bibnamefont {Varas}}, \bibinfo {author} {\bibfnamefont
  {I.}~\bibnamefont {Theophilou}}, \bibinfo {author} {\bibfnamefont
  {N.}~\bibnamefont {Helbig}}, \bibinfo {author} {\bibfnamefont {M.~J.}\
  \bibnamefont {Verstraete}}, \bibinfo {author} {\bibfnamefont
  {L.}~\bibnamefont {Stella}}, \bibinfo {author} {\bibfnamefont
  {F.}~\bibnamefont {Nogueira}}, \bibinfo {author} {\bibfnamefont
  {A.}~\bibnamefont {Aspuru-Guzik}}, \bibinfo {author} {\bibfnamefont
  {A.}~\bibnamefont {Castro}}, \bibinfo {author} {\bibfnamefont {M.~A.~L.}\
  \bibnamefont {Marques}}, \ and\ \bibinfo {author} {\bibfnamefont
  {A.}~\bibnamefont {Rubio}},\ }\bibfield  {title} {\enquote {\bibinfo {title}
  {{Real-space grids and the Octopus code as tools for the development of new
  simulation approaches for electronic systems}},}\ }\href {\doibase
  10.1039/C5CP00351B} {\bibfield  {journal} {\bibinfo  {journal} {Phys. Chem.
  Chem. Phys.}\ }\textbf {\bibinfo {volume} {17}},\ \bibinfo {pages} {31371}
  (\bibinfo {year} {2015})}\BibitemShut {NoStop}%
\bibitem [{\citenamefont {Jensen}\ and\ \citenamefont
  {Wasserman}(2018)}]{Jensen_2018}%
  \BibitemOpen
  \bibfield  {author} {\bibinfo {author} {\bibfnamefont {D.~S.}\ \bibnamefont
  {Jensen}}\ and\ \bibinfo {author} {\bibfnamefont {A.}~\bibnamefont
  {Wasserman}},\ }\bibfield  {title} {\enquote {\bibinfo {title} {Numerical
  methods for the inverse problem of density functional theory},}\ }\href
  {\doibase 10.1002/qua.25425} {\bibfield  {journal} {\bibinfo  {journal} {Int.
  J. Quantum Chem.}\ }\textbf {\bibinfo {volume} {118}},\ \bibinfo {pages}
  {e25425} (\bibinfo {year} {2018})}\BibitemShut {NoStop}%
\bibitem [{\citenamefont {Clevert}\ \emph {et~al.}(2015)\citenamefont
  {Clevert}, \citenamefont {Unterthiner},\ and\ \citenamefont
  {Hochreiter}}]{ELU}%
  \BibitemOpen
  \bibfield  {author} {\bibinfo {author} {\bibfnamefont {D.-A.}\ \bibnamefont
  {Clevert}}, \bibinfo {author} {\bibfnamefont {T.}~\bibnamefont
  {Unterthiner}}, \ and\ \bibinfo {author} {\bibfnamefont {S.}~\bibnamefont
  {Hochreiter}},\ }\bibfield  {title} {\enquote {\bibinfo {title} {{Fast and
  accurate deep network learning by exponential linear units (elus)}},}\ }\href
  {https://arxiv.org/abs/1511.07289} {\bibfield  {journal} {\bibinfo  {journal}
  {arXiv:1511.07289}\ } (\bibinfo {year} {2015})}\BibitemShut {NoStop}%
\bibitem [{Ign(2018)}]{Ignite}%
  \BibitemOpen
  \href@noop {} {\enquote {\bibinfo {title} {Ignite},}\ }\bibinfo
  {howpublished} {\url{https://github.com/pytorch/ignite},} (\bibinfo {year}
  {2018})\BibitemShut {NoStop}%
\bibitem [{\citenamefont {Kingma}\ and\ \citenamefont
  {Ba}(2014)}]{kingma2014adam}%
  \BibitemOpen
  \bibfield  {author} {\bibinfo {author} {\bibfnamefont {D.~P.}\ \bibnamefont
  {Kingma}}\ and\ \bibinfo {author} {\bibfnamefont {J.}~\bibnamefont {Ba}},\
  }\bibfield  {title} {\enquote {\bibinfo {title} {Adam: A method for
  stochastic optimization},}\ }\href@noop {} {\bibfield  {journal} {\bibinfo
  {journal} {arXiv:1412.6980}\ } (\bibinfo {year} {2014})}\BibitemShut
  {NoStop}%
\bibitem [{\citenamefont {Masters}\ and\ \citenamefont
  {Luschi}(2018)}]{masters2018revisiting}%
  \BibitemOpen
  \bibfield  {author} {\bibinfo {author} {\bibfnamefont {D.}~\bibnamefont
  {Masters}}\ and\ \bibinfo {author} {\bibfnamefont {C.}~\bibnamefont
  {Luschi}},\ }\bibfield  {title} {\enquote {\bibinfo {title} {Revisiting small
  batch training for deep neural networks},}\ }\href
  {https://arxiv.org/abs/1804.07612} {\bibfield  {journal} {\bibinfo  {journal}
  {arXiv:1804.07612}\ } (\bibinfo {year} {2018})}\BibitemShut {NoStop}%
\bibitem [{\citenamefont {Goodfellow}\ \emph {et~al.}(2016)\citenamefont
  {Goodfellow}, \citenamefont {Bengio},\ and\ \citenamefont
  {Courville}}]{Goodfellow-et-al-2016}%
  \BibitemOpen
  \bibfield  {author} {\bibinfo {author} {\bibfnamefont {I.}~\bibnamefont
  {Goodfellow}}, \bibinfo {author} {\bibfnamefont {Y.}~\bibnamefont {Bengio}},
  \ and\ \bibinfo {author} {\bibfnamefont {A.}~\bibnamefont {Courville}},\
  }\href@noop {} {\emph {\bibinfo {title} {Deep Learning}}}\ (\bibinfo
  {publisher} {MIT Press},\ \bibinfo {year} {2016})\ \bibinfo {note}
  {\url{http://www.deeplearningbook.org}}\BibitemShut {NoStop}%
\bibitem [{\citenamefont {Helbig}\ \emph {et~al.}(2011)\citenamefont {Helbig},
  \citenamefont {Fuks}, \citenamefont {Casula}, \citenamefont {Verstraete},
  \citenamefont {Marques}, \citenamefont {Tokatly},\ and\ \citenamefont
  {Rubio}}]{PhysRevA.83.032503}%
  \BibitemOpen
  \bibfield  {author} {\bibinfo {author} {\bibfnamefont {N.}~\bibnamefont
  {Helbig}}, \bibinfo {author} {\bibfnamefont {J.~I.}\ \bibnamefont {Fuks}},
  \bibinfo {author} {\bibfnamefont {M.}~\bibnamefont {Casula}}, \bibinfo
  {author} {\bibfnamefont {M.~J.}\ \bibnamefont {Verstraete}}, \bibinfo
  {author} {\bibfnamefont {M.~A.~L.}\ \bibnamefont {Marques}}, \bibinfo
  {author} {\bibfnamefont {I.~V.}\ \bibnamefont {Tokatly}}, \ and\ \bibinfo
  {author} {\bibfnamefont {A.}~\bibnamefont {Rubio}},\ }\bibfield  {title}
  {\enquote {\bibinfo {title} {Density functional theory beyond the linear
  regime: Validating an adiabatic local density approximation},}\ }\href
  {\doibase 10.1103/PhysRevA.83.032503} {\bibfield  {journal} {\bibinfo
  {journal} {Phys. Rev. A}\ }\textbf {\bibinfo {volume} {83}},\ \bibinfo
  {pages} {032503} (\bibinfo {year} {2011})}\BibitemShut {NoStop}%
\bibitem [{\citenamefont {Yu}\ \emph {et~al.}(2016{\natexlab{a}})\citenamefont
  {Yu}, \citenamefont {He},\ and\ \citenamefont
  {Truhlar}}]{doi:10.1021/acs.jctc.5b01082}%
  \BibitemOpen
  \bibfield  {author} {\bibinfo {author} {\bibfnamefont {H.~S.}\ \bibnamefont
  {Yu}}, \bibinfo {author} {\bibfnamefont {X.}~\bibnamefont {He}}, \ and\
  \bibinfo {author} {\bibfnamefont {D.~G.}\ \bibnamefont {Truhlar}},\
  }\bibfield  {title} {\enquote {\bibinfo {title} {{MN15-L: A New Local
  Exchange-Correlation Functional for Kohn–Sham Density Functional Theory
  with Broad Accuracy for Atoms, Molecules, and Solids}},}\ }\href
  {https://doi.org/10.1021/acs.jctc.5b01082} {\bibfield  {journal} {\bibinfo
  {journal} {J. Chem. Theory Comput.}\ }\textbf {\bibinfo {volume} {12}},\
  \bibinfo {pages} {1280} (\bibinfo {year} {2016}{\natexlab{a}})}\BibitemShut
  {NoStop}%
\bibitem [{\citenamefont {Yu}\ \emph {et~al.}(2016{\natexlab{b}})\citenamefont
  {Yu}, \citenamefont {He}, \citenamefont {Li},\ and\ \citenamefont
  {Truhlar}}]{C6SC00705H}%
  \BibitemOpen
  \bibfield  {author} {\bibinfo {author} {\bibfnamefont {H.~S.}\ \bibnamefont
  {Yu}}, \bibinfo {author} {\bibfnamefont {X.}~\bibnamefont {He}}, \bibinfo
  {author} {\bibfnamefont {S.~L.}\ \bibnamefont {Li}}, \ and\ \bibinfo {author}
  {\bibfnamefont {D.~G.}\ \bibnamefont {Truhlar}},\ }\bibfield  {title}
  {\enquote {\bibinfo {title} {{MN15: A Kohn–Sham global-hybrid
  exchange–correlation density functional with broad accuracy for
  multi-reference and single-reference systems and noncovalent
  interactions}},}\ }\href {\doibase 10.1039/C6SC00705H} {\bibfield  {journal}
  {\bibinfo  {journal} {Chem. Sci.}\ }\textbf {\bibinfo {volume} {7}},\
  \bibinfo {pages} {5032} (\bibinfo {year} {2016}{\natexlab{b}})}\BibitemShut
  {NoStop}%
\bibitem [{\citenamefont {Mardirossian}\ and\ \citenamefont
  {Head-Gordon}(2016)}]{Mardirossian2016}%
  \BibitemOpen
  \bibfield  {author} {\bibinfo {author} {\bibfnamefont {N.}~\bibnamefont
  {Mardirossian}}\ and\ \bibinfo {author} {\bibfnamefont {M.}~\bibnamefont
  {Head-Gordon}},\ }\bibfield  {title} {\enquote {\bibinfo {title}
  {{$\omega$B97M-V: A combinatorially optimized, range-separated hybrid,
  me\-ta-{GGA} density functional with {VV}10 nonlocal correlation}},}\ }\href
  {\doibase 10.1063/1.4952647} {\bibfield  {journal} {\bibinfo  {journal} {J.
  Chem. Phys.}\ }\textbf {\bibinfo {volume} {144}},\ \bibinfo {pages} {214110}
  (\bibinfo {year} {2016})}\BibitemShut {NoStop}%
\bibitem [{\citenamefont {Hollingsworth}\ \emph {et~al.}(2018)\citenamefont
  {Hollingsworth}, \citenamefont {Baker},\ and\ \citenamefont
  {Burke}}]{Hollingsworth2018}%
  \BibitemOpen
  \bibfield  {author} {\bibinfo {author} {\bibfnamefont {J.}~\bibnamefont
  {Hollingsworth}}, \bibinfo {author} {\bibfnamefont {T.~E.}\ \bibnamefont
  {Baker}}, \ and\ \bibinfo {author} {\bibfnamefont {K.}~\bibnamefont
  {Burke}},\ }\bibfield  {title} {\enquote {\bibinfo {title} {Can exact
  conditions improve machine-learned density functionals?}}\ }\href {\doibase
  10.1063/1.5025668} {\bibfield  {journal} {\bibinfo  {journal} {J. Chem.
  Phys.}\ }\textbf {\bibinfo {volume} {148}},\ \bibinfo {pages} {241743}
  (\bibinfo {year} {2018})}\BibitemShut {NoStop}%
\end{thebibliography}%

\end{document}